%% file: sample701.tex
\documentclass[twocolumn]{aastex701}

\defcitealias{diemer_13_scalingrel}{DKM13}
\defcitealias{diemer_14}{DK14}
\defcitealias{diemer_15}{DK15}
\defcitealias{diemer_17_sparta}{D17}
\defcitealias{diemer_20_catalogs}{D20}
\defcitealias{kannan_22}{K22}
\defcitealias{garaldi_22}{G22}
\defcitealias{smith_22}{S22}
\defcitealias{nelson_19}{N19}
\defcitealias{rodriguezpuebla_16}{RP+16}

\usepackage{xspace}
\usepackage{multirow}
\usepackage{graphicx}

\input{\includedir/macros_general}


\begin{document}

\title{A universal model for the accretion rates and formation times of dark matter halos}

\author[orcid=0000-0001-7072-570X,sname='Bera']{Ankita Bera}
\altaffiliation{CTC Post-Doctoral Fellow}
\affiliation{Department of Astronomy, University of Maryland, College Park, MD 20742, USA}
\email[show]{\href{mailto:ankitabm@umd.edu}{ankitabm@umd.edu}}  

\author[orcid=0000-0000-0000-0002, sname='Diemer']{Benedikt Diemer} 
\affiliation{Department of Astronomy, University of Maryland, College Park, MD 20742, USA}
\email{diemer@umd.edu}

\begin{abstract}

The formation histories of halos set the baseline rate at which galaxies accrete gas over cosmic time. While a number of models describe these histories and their derivative, the mass accretion rate (MAR), a simple and universal formula has remained elusive. Here we measure the median MARs and half-mass formation times of halos in dark matter-only and hydrodynamical simulations, in extremely different cosmologies ($\Lambda$CDM and Einstein–de Sitter), and across a wide range of redshifts ($z = 0$--$14$). We confirm that MARs increase with mass and redshift, and that they are virtually identical in hydrodynamical and dark matter-only simulations. We show that MARs are accurately described by a universal six-parameter function of three physical variables: the peak height $\nu$, the slope of the linear power spectrum $n_{\rm eff}$, and the effective linear growth rate $\alpha_{\rm eff}$. A complementary two-parameter fit for the formation redshift improves on the function of Lacey \& Cole by fixing one parameter to its physical value and adding a dependence on $n_{\rm eff}$. Our model is broadly consistent with some prescriptions from the literature but provides a larger range and higher accuracy at high redshifts and low masses. Our fitting functions are implemented in the publicly available \textsc{Colossus} toolkit. 

\end{abstract}

\section{Introduction} \label{sec:intro}

Understanding how dark matter halos grow over cosmic time is central to cosmology and galaxy formation. In the cold dark matter framework, halos provide the gravitational potential wells in which baryons cool and form galaxies \citep{white_78, 2010gfe..book.....M}. Halo growth governs the hierarchical buildup of structure and sets the boundary conditions for galaxy evolution, as the accretion of gas closely tracks the dark matter accretion at the halo scale \citep{faucher_giguere_11, van_de_11, wetzel_15}. This connection underpins both semi-analytic models of galaxy formation \citep[e.g.,][]{white_91, cole_00, somerville_08, henriques_15} and empirical models such as \textsc{UniverseMachine} \citep{behroozi_19} or \textsc{Emerge} \citep{moster_18}, which directly link halo mass assembly to star formation rates.

Simulations show that halos grow through a combination of processes: major mergers with similarly massive halos, minor mergers with lower-mass satellites, and the smooth accretion of diffuse material or unresolved halos \citep[e.g.,][]{genel_08, fakhouri_10}. Given the complexities of cosmic web formation, it is not clear that there should be a simple way to describe halo growth histories. Nonetheless, they can be approximated with semi-analytical techniques such as the extended Press-Schechter (EPS) formalism \citep{press_74, bond_91}, which creates random realizations of merger trees \citep{lacey_93, lacey_94, somerville_kolatt_99, neistein_08}. 

To simplify complicated merger trees, whether from simulations or EPS, halo growth is often described via the evolution of the most massive progenitor over  redshift, yielding so-called mass accretion histories (MAHs) \citep{lacey_93}. These MAHs can then be used to explore correlations with other halo properties such as environment, concentration, substructure fraction, spin, and the relative importance of major versus minor mergers \citep[see][and references therein]{gao_05_ab, wechsler_06, dalal_08_ab, wechsler_02, zhao_03_concentration, ludlow_13, vitvitska_02, maccio_08, fakhouri_09_environment}. Early studies established that halo MAHs can be described by simple functional forms, for example, a one-parameter exponential fit \citep{wechsler_02} or multi-parameter forms that provide improved fits \citep{vandenbosch_02_mah, tasitsiomi_04_clusterprof, fakhouri_10, mcbride_09}. 

All of these functions rely on some version of a formation time, for example, the redshift where a halo had acquired half its mass. It is difficult to predict this property of halos directly in the EPS framework \citep{lacey_93}, but it is easily measured in simulations \citep[e.g.,][]{navarro_97}. Formation time is a profound variable in the sense that MAHs exhibit some degree of universality across cosmological models when expressed in scaled variables \citep{zhao_09_acchist, vandenbosch_14}. Moreover, formation connects to other halo properties in profound ways. For example, concentration correlates strongly with formation redshift \citep{wechsler_02, zhao_03_mah, zhao_03_concentration} and environment influences assembly history with halos in denser regions forming earlier and growing more through major mergers \citep{maulbetsch_07}. 

Another important aspect of MAHs is their derivative, the mass accretion rate (MAR). Besides its tight connection to galaxy growth, the MAR (here defined as the logarithmic growth rate over a dynamical time) has been shown to profoundly affect the properties of halos. For example, higher MARs lead to a contraction of the density profile of halos at fixed mass \citep{diemer_14, adhikari_14, diemer_22}. Observationally, different MARs manifest as profoundly different populations of relaxed and unrelaxed galaxy clusters \citep[e.g.,][]{gouin_21}.

For all of these reasons, it is highly desirable to construct simple, universal fitting functions for MARs and formation times, which can then be used to approximate entire MAHs. Previously proposed fitting functions for MARs have often worked in units of halo mass, which leads to formulae that cannot be universal across cosmologies \citep[e.g.,][]{mcbride_09}. More physical models for MAHs are not always easy to evaluate due to complex, implicit functional forms \citep[e.g.,][]{vandenbosch_14}. MARs can of course also be derived from simpler MAH prescriptions, but those will only be as accurate as the input formation times. Most previous models for formation times were ultimately based on the formalism of \citet{lacey_93}, which is elegant but does not quite work when compared to simulations, especially if the mass fraction $f$ in their model is actually set to $1/2$ \citep[e.g.,][]{vandenbosch_02_mah, giocoli_07}. In summary, it is not clear that any current model accurately describes formation times across cosmology, redshift, and halo mass.

In this paper we construct fitting functions for the dimensionless dynamical MAR and formation time as a function of peak height $\nu$, effective power spectrum slope $n_{\rm eff}$, and effective growth rate of structure $\alpha_{\rm eff}$. This parameterization is based on the idea that structure in self-similar universes, meaning Einstein-de Sitter cosmologies with a power-law power spectrum, can only depend on the slope of the power spectrum, the expansion history (which one can artificially vary from its FLRW prediction), and the peak height (statistical rarity) of halos \citep{efstathiou_88, knollmann_08, joyce_21, leroy_21}. $\Lambda$CDM can be seen as an interpolation between Einstein-de Sitter universes with different power spectrum slopes. Within this framework, halo concentration can be described with simple functions \citep{diemer_15, diemer_19}. We similarly parameterize the formation time relative to some later epoch (which may or may not be $z = 0$), demonstrating that the simple formula of \citet{lacey_93} can be resurrected if we fix $f$ to its physical value of $1/2$ and add a dependence on $n_{\rm eff}$.

The paper is organized as follows. In Section~\ref{sec:method}, we describe the simulation suites and our analysis pipeline, including merger tree processing, the definition of the dynamical mass accretion rate, and resolution cuts. In Section~\ref{sec:res}, we present the median $\Gamma_{\rm dyn}$ as a function of peak height across cosmologies and simulation suites, introduce our universal fitting function for the MAR, and present a complementary fit for the half-mass formation redshift $z_{1/2}$. In section~\ref{sec:comp}, we compare our results with widely used MAR and MAH models in the literature. Finally, we summarize our conclusions and discuss implications in Section~\ref{sec:conclude}.

\section{Methods} \label{sec:method}

\subsection{Numerical Simulations} \label{subsec:sim}
%
\renewcommand{\arraystretch}{1.2}
\begin{deluxetable*}{clcccccccccccl}
\tabletypesize{\scriptsize}
\setlength{\tabcolsep}{3pt}  
\tablecaption{Details of the simulations }
\label{table:sims}
\tablehead{
\colhead{Simulation} & \colhead{Name} & \colhead{$L$} & \colhead{$N_{\rm DM,gas}^3$} & \colhead{$m_{\rm DM}$} & \colhead{$m_{\rm baryon}$} & \colhead{$\epsilon$} & \colhead{$z_{\rm initial}$} & \colhead{$z_{\rm final}$} & \colhead{$z_{\rm f-snap}$} & \colhead{Cosmology} & \colhead{Reference} \\
\colhead{suite} & \colhead{} & \colhead{$(\mpch)$} & \colhead{} & \colhead{$(\msunh)$} & \colhead{$(\msunh)$} & \colhead{$(\kpch)$} & \colhead{} & \colhead{} & \colhead{} & \colhead{} & \colhead{} & \colhead{}
}
\startdata
\multirow{15}{*}{\rotatebox[origin=c]{0}{\textbf{Erebos}}}
& L0016-WMAP7   & 16   & $1024^3$ & $2.86 \times 10^{5}$  & - & 0.25 & 99  & 4  & 24 & WMAP7  & this work \\
& L0031-WMAP7   & 31.25& $1024^3$ & $2.1 \times 10^{6}$  & - & 0.25  & 49  & 2  & 20 & WMAP7  & \citetalias{diemer_15} \\
& L0063-WMAP7   & 62.5 & $1024^3$ & $1.7 \times 10^{7}$  & - & 1.0   & 49  & 0  & 20 & WMAP7  & \citetalias{diemer_14} \\
& L0125-WMAP7   & 125  & $1024^3$ & $1.4 \times 10^{8}$  & - & 2.4   & 49  & 0  & 20 & WMAP7  & \citetalias{diemer_14} \\
& L0250-WMAP7   & 250  & $1024^3$ & $1.1 \times 10^{9}$  & - & 5.8   & 49  & 0  & 20 & WMAP7  & \citetalias{diemer_14} \\
& L0500-WMAP7   & 500  & $1024^3$ & $8.7 \times 10^{9}$  & - & 14    & 49  & 0  & 20 & WMAP7  & \citetalias{diemer_14} \\
& L1000-WMAP7   & 1000 & $1024^3$ & $7.0 \times 10^{10}$ & - & 33    & 49  & 0  & 20 & WMAP7  & \citetalias{diemer_13_scalingrel} \\
& L2000-WMAP7   & 2000 & $1024^3$ & $5.6 \times 10^{11}$ & - & 65    & 49  & 0  & 20 & WMAP7  & \citetalias{diemer_15} \\
& L0125-Planck  & 125 & $1024^3$ & $1.6 \times 10^{8}$  & - & 2.4    & 49  & 0  & 20 & Planck & \citetalias{diemer_15} \\
& L0250-Planck  & 250 & $1024^3$ & $1.3 \times 10^{9}$  & - & 5.8    & 49  & 0  & 20 & Planck & \citetalias{diemer_15} \\
& L0500-Planck  & 500 & $1024^3$ & $1.0 \times 10^{10}$ & - & 14     & 49  & 0  & 20 & Planck & \citetalias{diemer_15} \\
& L0100-PL-1.0  & 100 & $1024^3$ & $2.6 \times 10^{8}$  & - & 0.5    & 119 & 2  & 20 & PL, $n=-1.0$ & \citetalias{diemer_15} \\
& L0100-PL-1.5  & 100 & $1024^3$  & $2.6 \times 10^{8}$  & - & 0.5    & 99  & 1  & 20 & PL, $n=-1.5$ & \citetalias{diemer_15} \\
& L0100-PL-2.0  & 100 & $1024^3$  & $2.6 \times 10^{8}$  & - & 1.0    & 49  & 0.5 & 20 & PL, $n=-2.0$ & \citetalias{diemer_15} \\
& L0100-PL-2.5  & 100 & $1024^3$  & $2.6 \times 10^{8}$  & - & 1.0    & 49  & 0  & 20 & PL, $n=-2.5$ & \citetalias{diemer_15} \\
\hline
\multirow{6}{*}{\rotatebox[origin=c]{0}{\shortstack{\textbf{Illustris}\\\textbf{-TNG}}}}
& TNG50-1-Dark  & 35 & $2160^3$   & $3.7 \times 10^{5}$  & - & 2.2     & 99  & 0  & 20 & Planck & \citetalias{nelson_19} \\
& TNG100-1-Dark & 75 & $1820^3$   & $6.0 \times 10^{6}$    & - & 2.2   & 99  & 0  & 20 & Planck & \citetalias{nelson_19} \\
& TNG300-1-Dark & 205 & $2500^3$  & $7.0 \times 10^{7}$   & - & 2.2    & 99  & 0  & 20 & Planck & \citetalias{nelson_19} \\
& TNG50-1       & 35 & $2160^3$   & $3.1 \times 10^{5}$   & $5.7 \times 10^{4}$ & 2.2   & 99  & 0  & 20 & Planck & \citetalias{nelson_19} \\
& TNG100-1      & 75 & $1820^3$ & $5.1 \times 10^{6}$    & $9.4 \times 10^{5}$ & 2.2   & 99  & 0  & 20 & Planck & \citetalias{nelson_19} \\
& TNG300-1      & 205 & $2500^3$ & $4.0 \times 10^{7}$   & $7.6 \times 10^{6}$ & 2.2   & 99  & 0  & 20 & Planck & \citetalias{nelson_19} \\
\hline
\multirow{2}{*}{\rotatebox[origin=c]{0}{\textbf{Thesan}}}
& THESAN-Dark-1 & 64.7 & $2100^3$ & $5.46 \times 10^{6}$ & - & 1.5   & 49  & 0  & 20 & Planck & \citetalias{kannan_22, smith_22, garaldi_22} \\
& THESAN-1      & 64.7 & $2100^3$ & $4.6 \times 10^{6}$  & $8.6 \times 10^{5}$ & 1.5   & 49  & 5.5  & 20 & Planck & \citetalias{kannan_22, smith_22, garaldi_22} \\
\enddata
\tablecomments{$L$ denotes the box size in comoving units, $N^3$ the number of particles, $m_{\rm DM}$ the particle mass, and $\epsilon$ the force softening length in comoving units. The redshift range of each simulation is determined by the first and last redshifts $z_{\rm initial}$ and $z_{\rm final}$, but snapshots were output only between $z_{\rm f-snap}$ and $z_{\rm final}$. The cosmological parameters are given in Section~\ref{subsec:sim}. Here, `PL' indicates self-similar cosmologies with a power-law initial spectrum with slope $n$. The references correspond to \citet[][\citetalias{diemer_13_scalingrel}]{diemer_13_scalingrel}, \citet[][\citetalias{diemer_14}]{diemer_14}, \citet[][\citetalias{diemer_15}]{diemer_15}, \citet[][\citetalias{nelson_19}]{nelson_19}, \citet[][\citetalias{kannan_22}]{kannan_22}, \citet[][\citetalias{smith_22}]{smith_22}, and \citet[][\citetalias{garaldi_22}]{garaldi_22}. Our system for choosing force resolutions is discussed in \citetalias{diemer_14}.}
\end{deluxetable*}

To characterize the mass accretion rates across a wide range of halo masses, redshifts, and cosmologies, we require simulation suites that span dark matter-only (DMO) and hydrodynamic runs, multiple cosmologies, and sufficient dynamic range in both mass and box size. We use three complementary suites, namely \erebos, \itng, and \thesan, each of which contributes distinct advantages.

\begin{figure*}
    \centering
    \includegraphics[width=\textwidth]{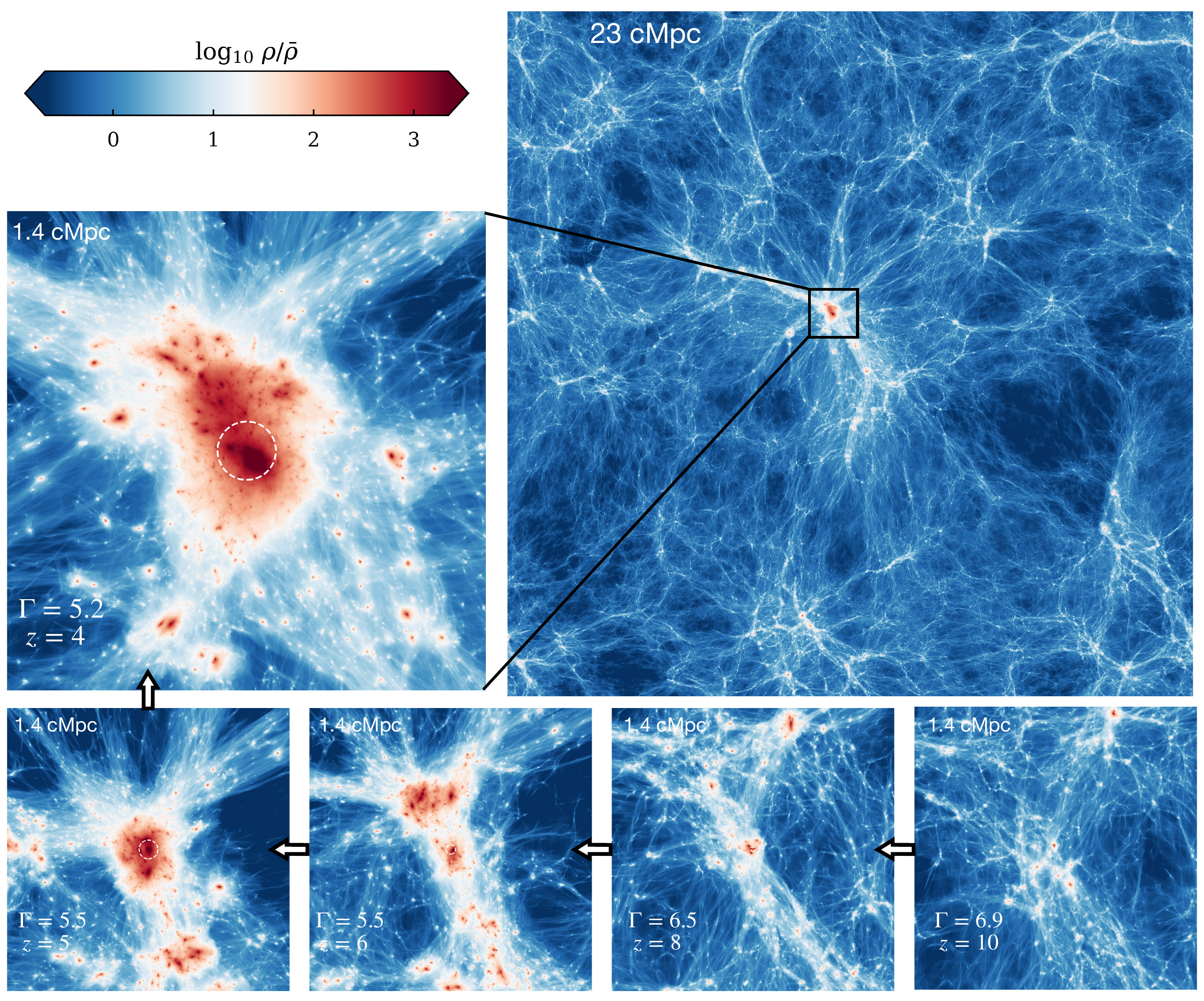}
    \caption{The dark matter density field from the L0016-WMAP7 simulation of the \erebos\ suite, illustrating the mass accretion of a single halo across cosmic time. The large right panel shows a projection through about 10\% of the simulation volume ($16\,\mpch\approx 23\,$cMpc) at $z=4$; the dashed box marks the region around a massive halo sitting at the intersection of several cosmic-web filaments. The remaining panels show $1.4\,$cMpc zoom-ins on the main progenitor of this halo, traced back in time from $z=4$ (upper left) through $z = 5, 6, 8,$ and $10$ (arrows indicate the tracking sequence), with the dynamical mass accretion rate $\Gamma_{\rm dyn}$ labelled in each. The dashed circle marks $R_{\rm 200m}$. Over this interval the halo grows vigorously, sustaining $\Gamma_{\rm dyn}\approx5$--$7$ through the accretion of material along filaments and through mergers.
    }
    \label{fig:illustration}
\end{figure*}

The \erebos suite of dissipationless $N$-body simulations \citep[][and this work]{diemer_14, diemer_15} provides a large range of box sizes (and thus peak heights), as well as self-similar cosmologies. Each of the suite's 15 simulations comprises $1024^3$ dark matter particles (see Table~\ref{table:sims}). Box sizes vary from $16\, \mpch $ to $ 2000\, \mpch$ (with corresponding mass resolutions), and the suite covers two $\Lambda$CDM cosmologies and four self-similar Einstein-de Sitter models with power-law initial power spectra of slopes $n= -1, -1.5, -2, -2.5$. The smallest box, $16\,\mpch$, was run specifically for this project and is designed to resolve low-mass halos at high redshift. Figure~\ref{fig:illustration} illustrates the resolving power of this box, showing the growth of a single high-redshift halo and its progenitors across $z=4$--$10$ through accretion along the cosmic web and mergers. The first $\Lambda$CDM cosmology matches that of the Bolshoi simulation \citep{klypin_11}, which is consistent with WMAP7 \citep{komatsu_11}. It assumes a flat $\Lambda$CDM universe with $\Omega_{\mathrm{m}} = 0.27$, $\Omega_{\mathrm{b}} = 0.0469$, $h = 0.70$, $\sigma_8 = 0.82$, and $n_s = 0.95$. The second is a Planck-like cosmology \citep{2014A&A...571A..16P} defined by $\Omega_{\mathrm{m}} = 0.32$, $\Omega_{\mathrm{b}} = 0.0491$, $h = 0.67$, $\sigma_8 = 0.834$, and $n_s = 0.9624$. The power spectra for the $\Lambda$CDM models were generated using \textsc{Camb} \citep{lewis_00}, and the initial conditions were created using \textsc{2LPTic} \citep{crocce_06}. All \erebos simulations were run with \textsc{Gadget-2} \citep{springel_05}.

The \itng suite adds hydrodynamic runs with full galaxy formation physics, allowing us to assess whether baryonic processes affect MARs and formation times relative to their DMO counterparts. The \itng simulation suite \citep{springel_18, pillepich_18, nelson_18, naiman_18, marinacci_18} is a set of gravity+magnetohydrodynamic cosmological simulations run with the moving-mesh code \textsc{Arepo} \citep{springel_10}. The suite comprises three flagship volumes with box sizes of approximately 50, 100, and 300 comoving Mpc (TNG50, TNG100, and TNG300 respectively), each run at multiple resolution levels. For this work, we use the highest-resolution realizations: TNG50-1, TNG100-1, and TNG300-1 (see Table~\ref{table:sims} for details). We analyze both the full hydrodynamical runs and their dark-matter-only counterparts (TNG50-1-Dark, TNG100-1-Dark, TNG300-1-Dark), which share identical initial conditions and numerical parameters. The simulations adopt \citet{2014A&A...571A..16P} cosmological parameters: $\Omega_{\mathrm{m}} = 0.3089$, $\Omega_{\Lambda} = 0.6911$, $\Omega_{\mathrm{b}} = 0.0486$, $h = 0.6774$, $\sigma_8 = 0.8159$, and $n_s = 0.9667$. Initial conditions were generated at $z = 127$ using the Zel'dovich approximation. The TNG galaxy formation model includes comprehensive treatments of gas cooling, star formation, stellar evolution and chemical enrichment, supernova-driven galactic winds, supermassive black hole seeding, growth and multi-mode feedback, as well as magnetic field amplification \citep{weinberger_17, pillepich_18}. 

Finally, \thesan \citep{kannan_22, smith_22, garaldi_22} extends our sample to the epoch of reionization with radiation-magnetohydrodynamic simulations designed to simultaneously capture the large-scale statistical properties of the intergalactic medium (IGM) during reionization and the detailed characteristics of the galaxies driving it. \thesan allows us to probe MARs at $z \gtrsim 6$ where standard hydrodynamic simulations may not model the reionization environment accurately (although this turns out to be unimportant). The cosmology and baryonic models are the same as for \itng, but \thesan was run with \textsc{Arepo-RT}, which adds radiative transfer to \textsc{Arepo}. 

\subsection{Merger tree analysis} \label{subsec:analysis}

We restrict our sample to host halos and trace their evolution along the main progenitor branch of the merger tree. While this is the most commonly used definition of the MAH, we note that one can also sum over all progenitors at each redshift \citep[e.g.,][]{ludlow_16}. One potential difficulty is that our simulation suites use different halo finders and merger tree codes. Specifically, halos in \erebos are identified using \textsc{Rockstar} \citep{behroozi_13_rockstar} and converted to merger trees using \textsc{consistent-trees} \citep{behroozi_13_trees}. We use augmented trees that already contain mass accretion rates \citep{diemer_20_moria}. In contrast, halos in  \itng and \thesan are found with \textsc{Subfind} \citep{springel_01, dolag_09}. Both halo finders use Friends-of-Friends (FoF) algorithms, but \textsc{Rockstar} does so in phase space and \textsc{Subfind} in configuration space. We use merger trees constructed with \textsc{SubLink} \citep{rodriguez-gomez_15} and \textsc{LHaloTree} \citep{springel_05_millennium} for \itng and \thesan, respectively. 

The different approaches could in principle lead to systematic differences in halo masses, merger rates, and tree statistics, but comparisons of halo-finding algorithms have shown good agreement for host halos \citep{knebe_11}, with the choice of halo finder primarily influencing merger tree statistics such as tree length and inferred merger rates \citep{avila_14}. We show in Section~\ref{sec:res} that the methodological differences have almost no impact on MARs.


\subsection{Peak height and mass accretion rate} \label{subsec:mass_acc}
We bin halos by their peak height, $\nu$, rather than by mass, since halo properties are expected to be approximately redshift-independent at fixed $\nu$ \citep{lacey_94, sheth_99}. The peak height is defined as
\begin{equation}
\nu \equiv \frac{\delta_c}{\sigma(M,z)} = \frac{\delta_c}{\sigma(M,z=0) \, D_+(z)},
\end{equation}
where, $\delta_c = 1.686$ is the critical overdensity for collapse derived from the spherical top-hat model \citep{gunn_72}, and $D_+(z)$ is the linear growth factor of density fluctuations normalized to unity at $z = 0$. We neglect the weak dependence of $\delta_c$ on cosmology and redshift \citep{mo_96}. The function $\sigma(M,z)$ is the root-mean-square fluctuation of the linear density field smoothed over a scale $R$, and is given by
\begin{equation}
\sigma^2(R) = \frac{1}{2\pi^2} \int_0^{\infty} k^2 P(k) \left| \widetilde{W}(kR) \right|^2 \, dk,
\end{equation}
where, $P(k)$ is the linear matter power spectrum and $\widetilde{W}(kR)$ is the Fourier transform of the spherical top-hat window function. We adopt the analytic fitting formula for $P(k)$ from \citet{eisenstein_98}, normalized to yield $\sigma(8\,h^{-1}\mathrm{Mpc}) = \sigma_8$ appropriate to the cosmology of each simulation. The mass variance is expressed as $\sigma(M) = \sigma(R[M])$, where mass and radius are related by $M = (4\pi/3)\,\rho_m(z=0)\,R^3$. For our analysis, we compute $\nu$ using $M = M_{\mathrm{200m}}$. Binning halos uniformly in $\nu$ tends to emphasize higher-mass halos, especially at low redshift. We extract halo masses and peak heights at a series of discrete redshifts spanning $z = 0$ to $z = 12$ for \erebos, $z = 10$ for \itng, and $z = 14$ for \thesan, yielding a combined sample of over $12$ million halo accretion measurements. The upper redshift limit for each suite is set by the resolution criteria described later in this section. 

At each redshift, we define the accretion rate of a halo based on its growth over one dynamical time, following the definition of \citet{diemer_17_sparta},
\begin{equation}
    \Gamma_{\rm dyn} (t) = \frac{\log M(t) - \log M(t - t_{\rm dyn})}{\log a(t) - \log a(t - t_{\rm dyn})} 
    \label{eq:gamma_def}
\end{equation}
where $a$ is the cosmic scale factor and $M$ is the halo mass. We again choose $M = M_{\mathrm{200m}}$. The dynamical time is defined as the crossing time,
\begin{equation}
    t_{\rm dyn} (z) \equiv t_{\rm cross} (z) = \frac{2R_{\rm 200m}}{V_{\rm 200m}} ,
    \label{eq:t_dyn}
\end{equation}
where $V_{\rm 200m} = \sqrt{G M_{\rm 200m}/R_{\rm 200m}}$. The crossing time has been shown to correspond to the timescale over which halo structure is most sensitive to the MAR \citep{shin_23}. For halos whose main progenitor was not a host halo one dynamical time earlier, typically because it had become a subhalo of a more massive system, we instead measure the MAR at the closest available snapshot that lies between half or twice $t_{\rm dyn}$, if possible \citep{diemer_20_moria}. In practice, this situation is relatively rare and has a negligible impact on our results. While the specific choice of overdensity ($M = M_{\mathrm{200m}}$) also has little impact on the inferred MARs \citep{xhakaj_19}, the matter flow through $R_{\rm 200m}$ is less likely to be affected by baryonic processes compared to radii such as $R_{\rm 200c}$. However, we note that accretion shocks can extend well beyond $R_{\rm 200m}$ \citep{sen_26}, so that even this choice does not fully isolate the MAR from baryonic physics. 


We impose several criteria to ensure that only well-resolved, physically distinct halos are included in our analysis. First, we discard all satellite because their MARs are generally not well defined. Second, we require halos to contain at least 50 DM particles at both the initial and final snapshots to ensure reasonably well-measured overdensity masses \citep{benson_17}. This threshold introduces a subtle but important bias: halos that have undergone very rapid accretion had correspondingly smaller progenitors, which are more likely to fall below the resolution limit. As a result, the highest accretion rates at any given final mass are preferentially lost from the sample. At fixed resolution, the severity of this bias depends on halo mass. At high $\nu$, the maximum accretion rate permitted by the resolution is far larger than any physically plausible value. At low $\nu$, however, this maximum permitted rate decreases and begins to encroach on the tail of the distribution, progressively removing halos with the highest MARs and biasing the median downward. We therefore identify the peak height $\nu_{\rm min}(z)$ below which this bias becomes significant as the peak height where MARs that lie $3$--$\sigma$ above the mean cannot be resolved. We estimate this limit based on the fitting function of \citet{diemer_20_moria}, which was calibrated on a smaller set of simulations and serves as a rough approximation to the range of physically expected MARs. We limit the function to $z\leq 8$ since it was not calibrated at higher redshift. Excluding halos below $\nu_{\rm min}(z)$ guarantees that only the strongest outliers in the MAR distribution (above $3$--$\sigma$) are lost. We have verified that these outliers have a negligible effect on the median by varying the threshold between $2$--$\sigma$ and $5$--$\sigma$.

\begin{figure*}[!ht]
    \centering
    \includegraphics[trim = 2mm 14mm 2mm 2mm, clip, scale=0.7]{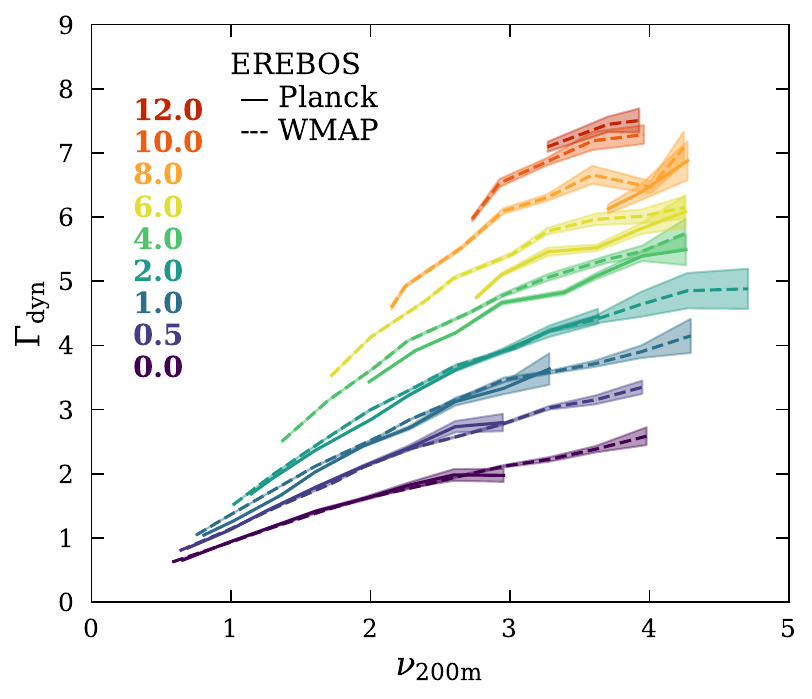}
    \includegraphics[trim = 13mm 14mm 2mm 2mm, clip, scale=0.7]{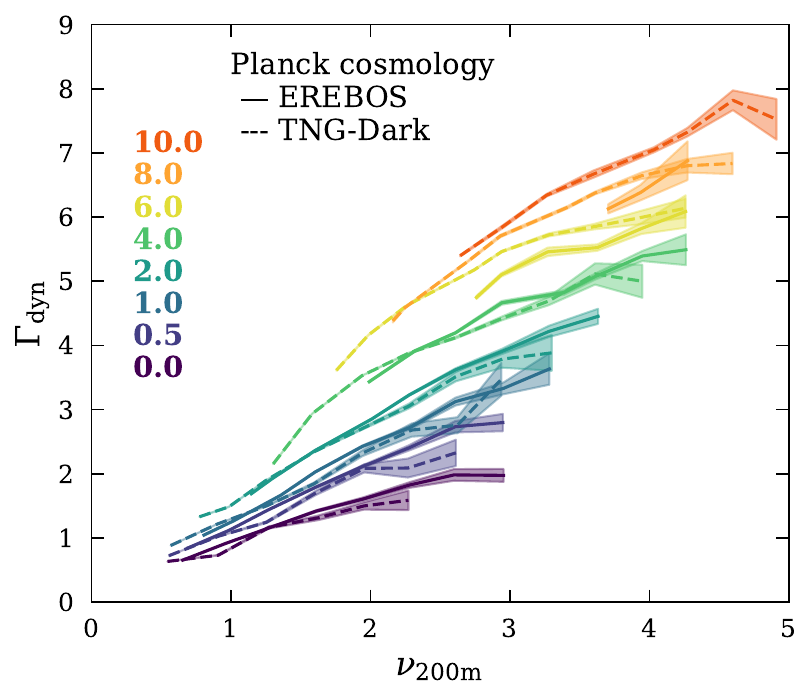}
    \includegraphics[trim = 2mm 2mm 2mm 2mm, clip, scale=0.7]{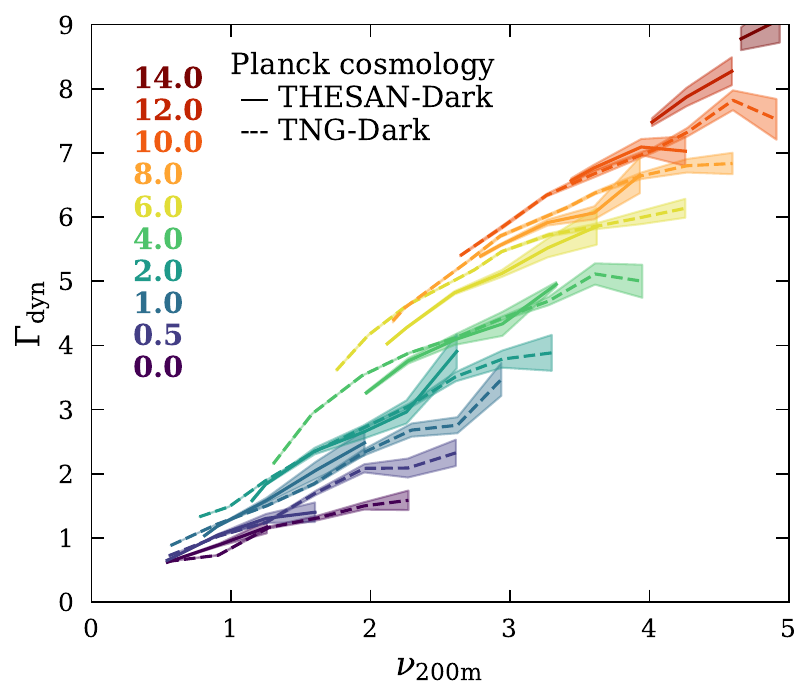}
    \includegraphics[trim = 13mm 2mm 2mm 2mm, clip, scale=0.7]{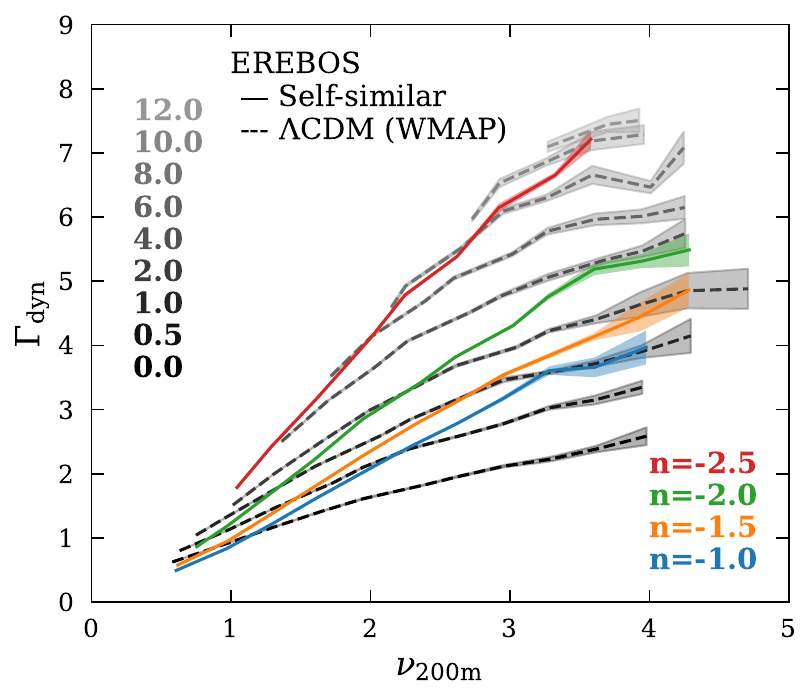}
    \caption{Dependence of the median mass accretion rate $\Gamma_{\rm dyn}$ on cosmology and simulation suite. Each panel shows $\Gamma_{\rm dyn}$ as a function of peak height $\nu_{200\rm m}$ at multiple redshifts (colors as labeled), with shaded regions indicating $1$--$\sigma$ bootstrap uncertainties. \textbf{Top left:} Comparison of Planck (solid) and WMAP7 (dashed) cosmologies using the \erebos suite. The two cosmologies yield similar results at low redshift but diverge at $z \gtrsim 4$. \textbf{Top right:} Comparison of \erebos (solid) and TNG (dashed) simulations with near-identical Planck cosmologies, illustrating simulation-dependent offsets arising from differences in resolution, halo finding, and merger tree construction. \textbf{Bottom left:} Comparison of \thesan-Dark (solid) and TNG-Dark (dashed) at Planck cosmology, highlighting the impact of different merger tree algorithms (\textsc{LHaloTree} vs.\ \textsc{SubLink}). \textbf{Bottom right:} Comparison of self-similar Einstein-de Sitter cosmologies (solid; spectral indices as labeled) with the WMAP7 $\Lambda$CDM cosmology (dashed). The power spectrum slope in $\Lambda$CDM becomes shallower with size scale, and thus with increasing $\nu$ and decreasing $z$, effectively interpolating between self-similar universes.}
    \label{fig:all_mar}
\end{figure*}

After applying the resolution cuts to each individual simulation, we aggregate results across all boxes to obtain the median accretion rate as a function of $\nu$. We require at least 60 halos per $\nu$ bin to avoid noisy medians. We quantify the uncertainty in the medians by performing bootstrap resampling with replacement, generating 1000 random samples per bin. In the following figures, the standard deviations of the bootstrap samples are shown as shaded contours.

\section{Results} \label{sec:res}

In this section we present how the median MARs depend on peak height, cosmology, redshift, and simulation details (Section~\ref{sec:res:med_mar}). We build universal models both for the medians (Section~\ref{sec:res:uni_rel}) and the distribution of MARs (Section~\ref{sec:res:distribution}). We then show that a similarly universal description can be applied to formation times (Section~\ref{sec:res:z_half}).

\subsection{Median mass accretion rates} \label{sec:res:med_mar}

Figure~\ref{fig:all_mar} shows the median mass accretion rate, $\Gamma_{\rm dyn}$, as a function of peak height for several simulation suites and cosmologies. The colored curves denote different redshifts, spanning $z=0$ to $z=14$ with $1$--$\sigma$ uncertainty shown as shaded regions. Across all redshifts, $\Gamma_{\rm dyn}$ increases monotonically with $\nu$. This trend indicates that rarer, higher-$\nu$ peaks experience systematically higher accretion activity when expressed in dynamical time units, consistent with the expectation that the most rapidly assembling objects at a given epoch occupy the high-$\nu$ tail of the population. At fixed $\nu$, $\Gamma_{\rm dyn}$ also increases with redshift (from purple to red). This captures the accelerated pace of structure growth at early times, where typical mass-doubling times are shorter relative to the dynamical time.

The top left panel highlights the differences in MAR derived from the merger trees of the \erebos simulation for the Planck (solid) and WMAP7 (dashed) cosmologies. At low redshift, the two relations are hard to distinguish but at high redshift, the MARs in the WMAP7 cosmology are noticeably higher at fixed $\nu$. 
This comparison illustrates that a MAR model calibrated on a single cosmology cannot be assumed to generalize to others, which motivates us to look for cosmology-dependent quantities that explain the deviation.

Besides cosmology, there are several other effects that could produce systematic offsets between simulation suites even at similar cosmology. First, cosmic variance means that small simulation boxes are not fair samples of the universe, so that the abundance and properties of halos at fixed $\nu$ can deviate from the true cosmic mean. This is a particularly strong factor for the smallest boxes in our sample ($L=16$--$32\,{\rm Mpc}\,h^{-1}$).
Second, different halo finders use different definitions of the halo center and define subhalos differently (e.g., via spherical overdensity or membership in FOF groups). Moreover, merger trees could differ due to different algorithms, snapshot spacing \citep{benson_12} and branch selection can change the inferred growth between snapshots, which can systematically affect the median $\Gamma_{\rm dyn}$ at fixed $\nu$.

To test these effects, the top right panel of Figure~\ref{fig:all_mar} compares $\Gamma_{\rm dyn}$ between the \erebos (solid) and \itng (dashed) simulation suites with slightly different Planck-like cosmologies. Despite the substantial methodological differences between the two suites (namely, different simulation codes, halo finders, subhalo definitions, merger-tree algorithms, mass and time resolutions, and box sizes), the inferred median MAR relations are remarkably congruent over the full range of $\nu$ and $z$. 
This finding confirms that halo finders and merger-tree algorithms broadly agree on the properties of isolated halos \citep{knebe_11, avila_14}. The cosmologies adopted in \erebos and \itng are both Planck-like but not exactly identical, meaning that small residual differences may also reflect the dependence on cosmology.
We further test the impact of methodology in the bottom left panel by comparing \thesan (solid) and \itng (dashed), which share the same galaxy formation model and similar cosmological parameters but differ in merger-tree construction (LHaloTree vs. SubLink).
The overall agreement is again encouraging.

To understand the origin of the slight differences between our (relatively similar) $\Lambda$CDM cosmologies, the bottom right panel of Figure~\ref{fig:all_mar} compares MARs in WMAP7 to those in the self-similar universes in the \erebos suite. For each of those simulations, we combine the relationships measured at different ``redshifts,'' which is not a physically meaningful quantity in a simulation without a physical mass or time scale. We have verified that the relations agree. Between simulations, the relations differ vastly at fixed $\nu$, demonstrating that the power spectrum slope strongly influences the median MAR. In $\Lambda$CDM, the slope depends on the halo scale, which changes with both peak height and redshift. Thus, the relation at any one redshift does not match any one self-similar universe, but matching them via an effective slope forms the backbone of our model in Section~\ref{sec:res:uni_rel}.

\begin{figure}
    \centering
    \includegraphics[width=0.45\textwidth]{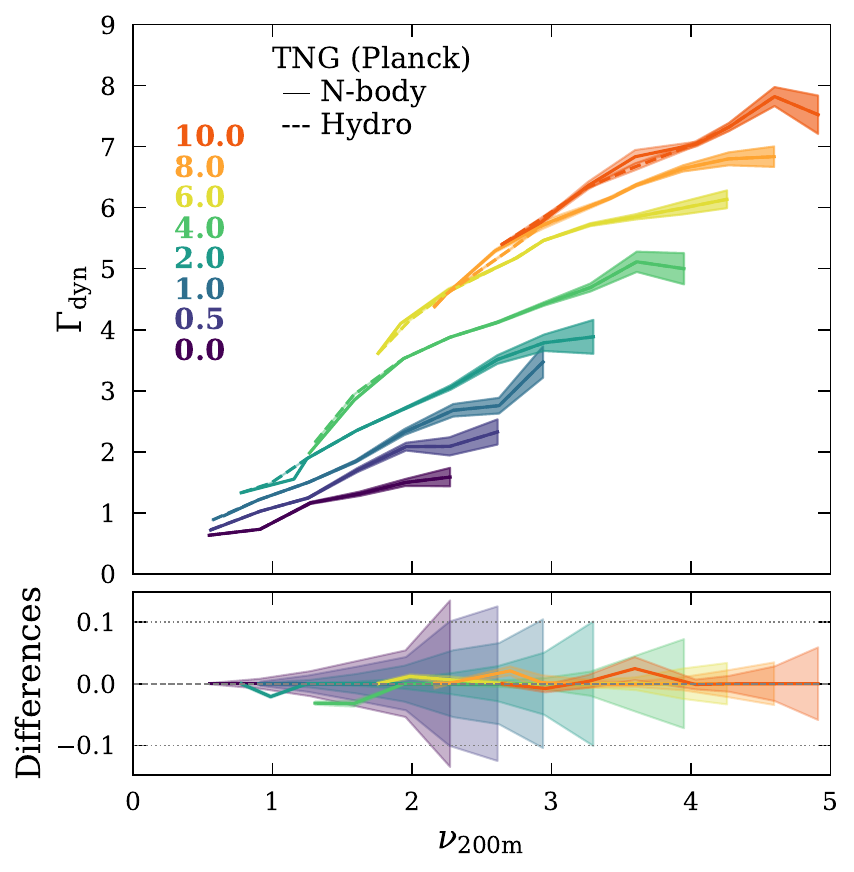}
    \caption{\textbf{Top panel:} Comparison of $\Gamma_{\rm dyn}$ for \itng dark-matter-only (solid) and hydrodynamical (dashed) simulations. \textbf{Bottom panel:} The fractional differences between dark-matter-only and hydrodynamical simulations $(\Gamma_{\rm hydro} - \Gamma_{\rm Dm})/ \Gamma_{\rm Dm}$. The excellent agreement ($\lesssim 0.1$) between N-body and Hydro runs demonstrates that baryonic processes have minimal effect on mass accretion rates for the halo mass range probed here.}
    \label{fig:tng_dm_hydro}
\end{figure}
\renewcommand{\arraystretch}{1.2}
\begin{deluxetable}{lcc}
\tablecaption{Best-fit values and $1$--$\sigma$ uncertainties on the parameters for the universal mass accretion rate, scatter, and half-mass formation redshift.\label{tab:fit_params}}
\tablehead{
\colhead{Parameters} \hspace{.9cm} & \colhead{Value} \hspace{.9cm} & \colhead{$\pm 1$--$\sigma$} \hspace{.9cm}
}
\startdata
\sidehead{Median MAR (Eq.~\ref{eq:fit_func}--\ref{eq:scatter})}
\hline
$a_0$ & $\phantom{-}1.143$ & $0.017$ \\
$a_1$ & $-0.002$           & $0.022$ \\
$b_0$ & $\phantom{-}1.212$ & $0.021$ \\
$b_1$ & $\phantom{-}1.830$ & $0.032$ \\
$c_0$ & $-1.439$           & $0.065$ \\
$c_1$ & $-0.506$           & $0.132$ \\
\hline
\sidehead{Scatter in MAR (Eq.~\ref{eq:scatter})}
\hline
$s_0$ & $0.680$ & $0.045$ \\
$s_1$ & $0.134$ & $0.056$ \\
$s_2$ & $0.497$ & $0.020$ \\
\hline
\sidehead{Formation redshift (Eq.~\ref{eq:C_of_neff})}
\hline
$C_0$ & $\phantom{-}0.646$ & $0.002$ \\
$C_1$ & $-0.062$ & $0.001$ \\
\enddata
\end{deluxetable}

While we have verified that the simulation and analysis algorithms have at most minor impacts on the MAR, we also wish to quantify any effects due to baryonic physics. In Figure~\ref{fig:tng_dm_hydro}, we compare the \itng hydrodynamical runs to their dark-matter-only counterparts (TNG-Dark) at fixed cosmology. The median $\Gamma_{\rm dyn}$ agrees to better than $\Delta\Gamma_{\rm dyn} \lesssim 0.05$, and the differences are well within the $1$--$\sigma$ uncertainty for almost all bins. The near-perfect overlap highlights that baryonic physics can only moderately alter the mass within $R_{\rm 200m}$, for example because hydrodynamical effects (such as accretion shocks) can prevent baryons from entering the halo at larger radii, or because gas is ejected by feedback. Moreover, $\Gamma_{\rm dyn}$ is defined as a logarithmic derivative, meaning that only changes in mass that vary with time are captured.


\subsection{Universal model for the MAR} \label{sec:res:uni_rel}

\begin{figure*}
    \centering
    \includegraphics[trim = 2mm 14mm 2mm 2mm, clip, scale=0.6]{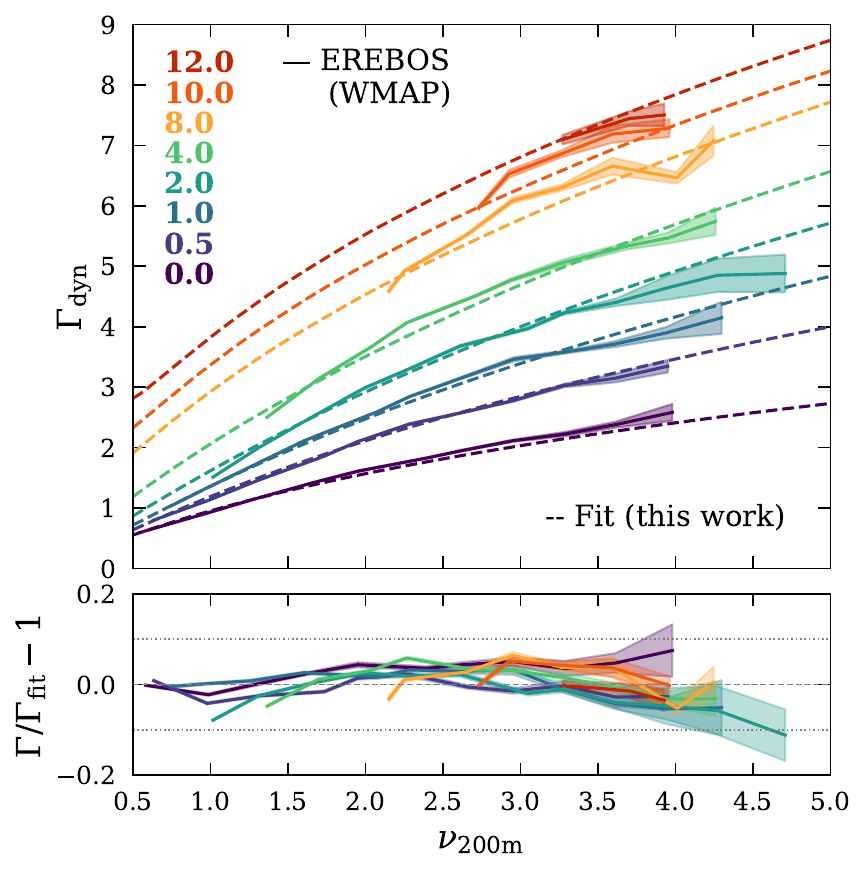}
    \includegraphics[trim = 19.4mm 14mm 2mm 2mm, clip, scale=0.6]{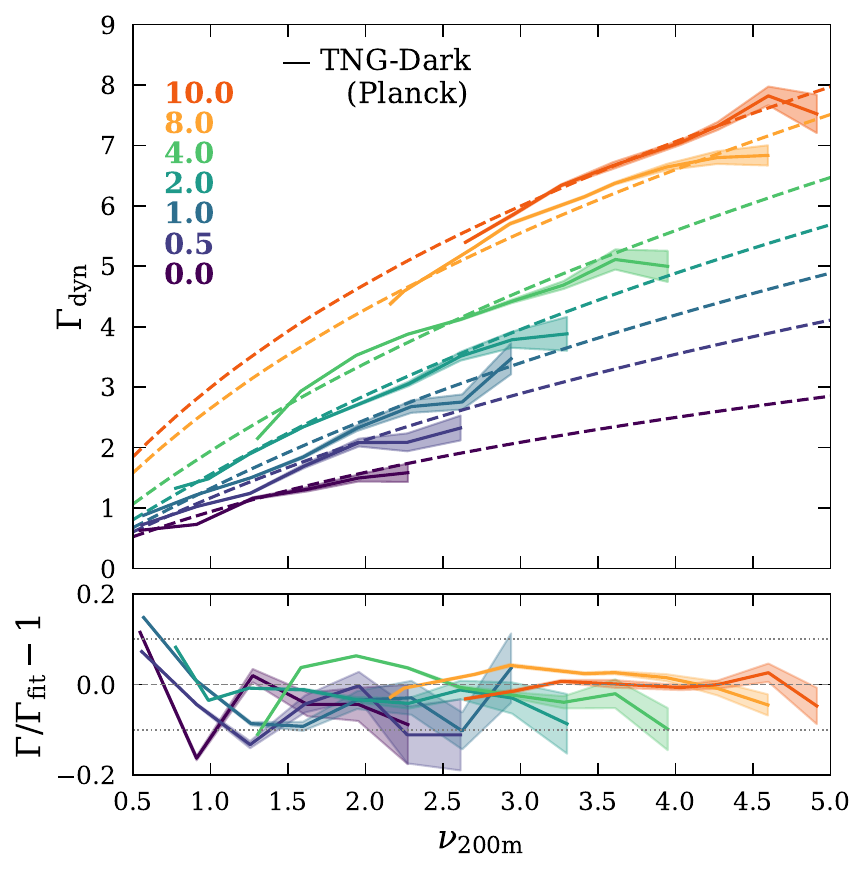}
    \includegraphics[trim = 2mm 2mm 2mm 2mm, clip, scale=0.6]{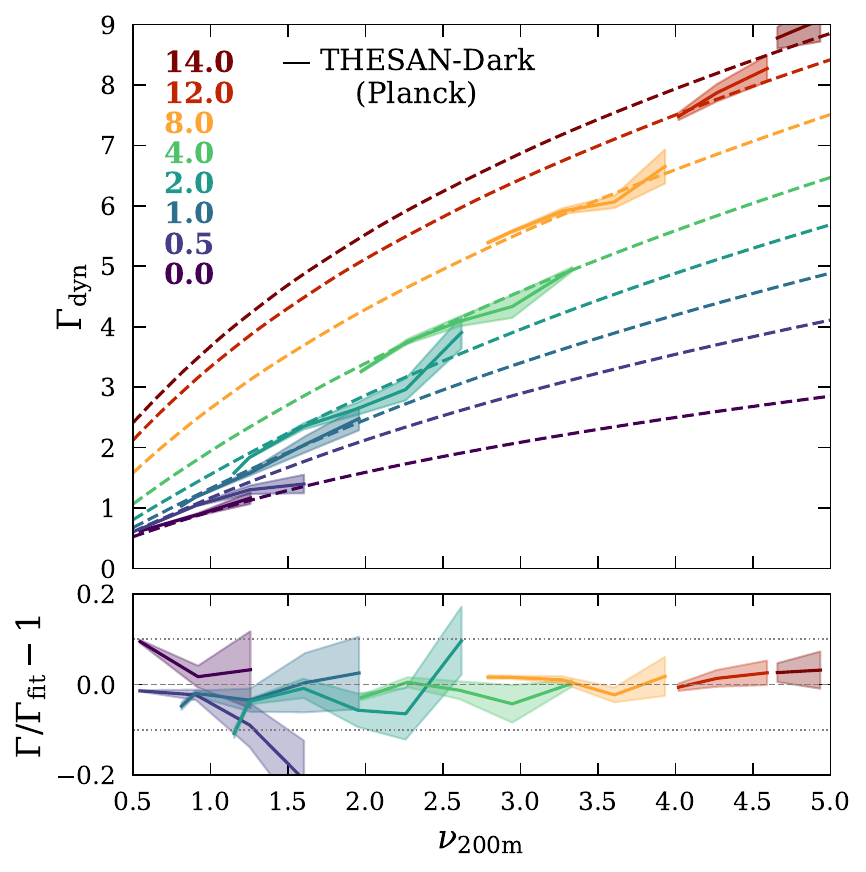}
    \includegraphics[trim = 19.4mm 2mm 2mm 2mm, clip, scale=0.6]{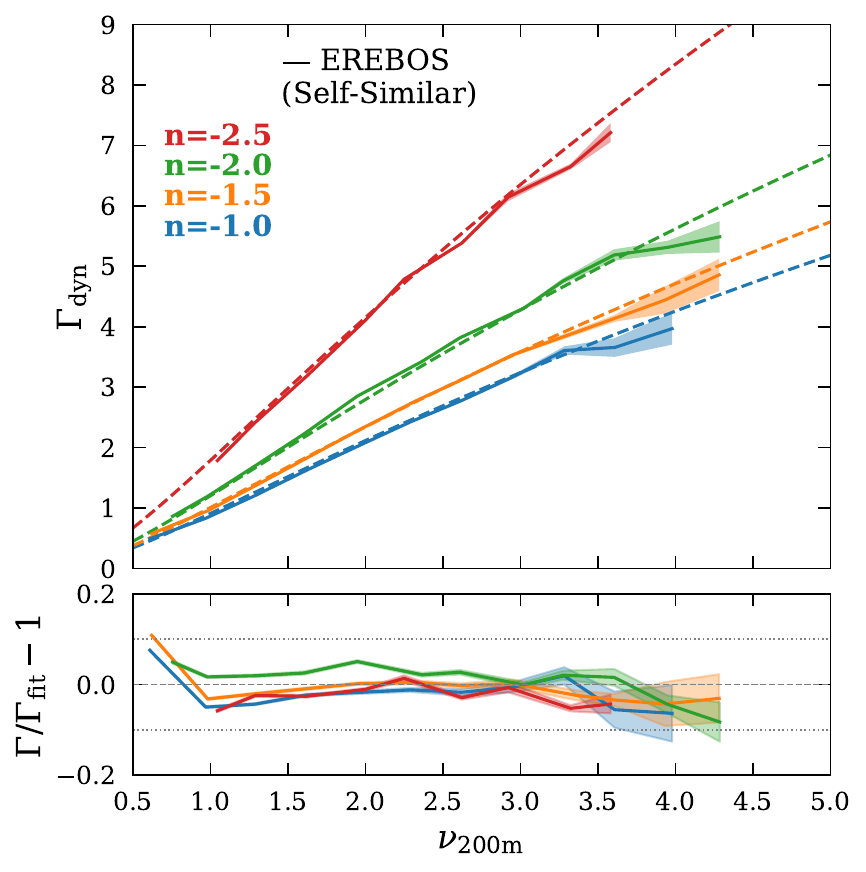}
    \caption{Validation of our universal fitting function across simulations, cosmologies, and redshifts. Each panel shows $\Gamma_{\rm dyn}$ as a function of peak height, with solid lines indicating the median values from simulations and dashed lines showing predictions from our six-parameter fitting function (Equation~\ref{eq:fit_func}). Colors denote redshift as labeled, with shaded regions representing $1$--$\sigma$ bootstrap uncertainties. The lower sub-panels show fractional residuals, with dotted horizontal lines marking the $\pm 10\%$ level. \textbf{Top left:} \erebos suite with WMAP7 cosmology, spanning $z = 0$ to $z = 12$. \textbf{Top right:} TNG-Dark suite with Planck cosmology, extending to $z = 10$. \textbf{Bottom left:} \thesan-Dark simulation with Planck cosmology, reaching $z = 14$. \textbf{Bottom right:} Self-similar Einstein-de Sitter cosmologies from \erebos with power-law initial spectra of slopes $n = -1.0$, $-1.5$, $-2.0$, and $-2.5$ (colors as labeled). The fitting function achieves better than 10\% accuracy across all $\Lambda$CDM simulations from $z = 0$ to $z = 14$ and reproduces the systematic variation with spectral index in the self-similar models, demonstrating that our parametrization in terms of $\nu$, $n_{\rm eff}$, and $\alpha_{\rm eff}$ captures the essential physics of halo mass assembly across a wide range of cosmological conditions.}
    \label{fig:all_fitting}
\end{figure*}
A central goal of this work is to express MARs in a form that is universal across redshift, mass, and cosmology. The theoretical basis for such a description is the self-similar character of gravitational collapse, which dictates that halo properties should depend only on their peak height, the shape of the power spectrum, and the expansion history of the universe. The power spectrum determines both the relative abundance of progenitors on different scales as well as the shape of the initial peaks, which directly translates into the accretion histories \citep{bardeen_86, dalal_10}. The linear growth rate of structure determines when and how fast perturbations collapse. We approximately describe these effects based on two physically motivated variables in addition to peak height, namely the scale-dependent effective slope of the power spectrum, $n_{\rm eff}$, and the effective growth rate of structure, $\alpha_{\rm eff}$, which depends only on redshift in a given cosmology. We define $n_{\rm eff}$ as in \citet{diemer_19}, 
\begin{equation}
    n_{\rm eff} \equiv -2 \frac{d\ln \sigma(R)}{d\ln R} \bigg|_{R=\kappa R_{\rm L}} - 3 \,,
    \label{eq:neff}
\end{equation}
where $R_{\rm L}$ is the Lagrangian radius of a halo and $\kappa$ is a free parameter that we fix to $\kappa = 1$ in the rest of our analysis because we find no strong preference for other values. In self-similar cosmologies we have $n_{\rm eff} = n$ by construction, but in $\Lambda$CDM it varies as a function of scale. Since we set that scale to the Lagrangian radius, $n_{\rm eff}$ varies with both halo mass and redshift: cluster-scale halos at low redshift correspond to relatively shallow effective slopes ($n_{\rm eff} \sim -1.5$), while dwarf-scale halos or high redshifts lead to steeper slopes ($n_{\rm eff} \lesssim -2.3$ \citep[e.g., Figure 4 of][]{diemer_15}. We emphasize that there is no one ``correct'' definition of the effective slope. For example, we could also define $n_{\rm eff} \equiv d\ln P(k)/d\ln k$ \citep{bardeen_86, jing_98, diemer_15}, but the former definition was found to more closely correlate with concentration \citep{diemer_19}. For any quantity we wish to parameterize, one can optimize a filter and scale over which to characterize the power spectrum \citep[as demonstrated for the Einasto shape parameter by][]{brown_20}.

Finally, the eﬀective linear growth rate $\alpha_{\rm eff}$ tracks departures from matter-dominated expansion: it equals unity in an Einstein-de Sitter universe and decreases at late times in $\Lambda$CDM as dark energy suppresses the growth of perturbations. We again use the same definition as \citet{diemer_19}, \footnote{Recently, \citet{sanchez_22} proposed an `evolution mapping' framework that relies on a somewhat more complex integral quantity to capture the expansion history dependence. One could in principle adopt that approach, but we find that $\alpha_{\rm eff}$ is sufficient for our purposes.}
\begin{equation}
    \alpha_{\rm eff} (z) \equiv - \frac{d \ln D(z)}{d \ln (1+z)} \, .
\end{equation}
Whether this parameterization is as effective for MARs as for concentrations is not guaranteed, and demonstrating this is one of the main results of this paper. After experimenting with a range of functional forms, including polynomials, power laws, and combinations thereof,\footnote{We also explored automated symbolic regression using \textsc{PySR} \citep{cranmer_pysr}, but the resulting expressions either lacked a clear physical interpretation or did not improve on the performance of the simple form adopted here.} we find that $\Gamma_{\rm dyn}$ is well described by 
\begin{equation}
    \Gamma(\nu_{\rm 200m}, n_{\rm eff}, \alpha_{\rm eff}) = A_0 \times [\ln(1+\nu_{\rm 200m})]^{A_1},
    \label{eq:fit_func}
\end{equation}
where the coefficients $A_\rmi \, (i=0, 1)$ are functions of $n_{\rm eff}$ and $\alpha_{\rm eff}$,
\begin{equation}
    A_\rmi = \left( a_\rmi + \frac{b_\rmi}{3 + n_{\rm eff}}  + c_\rmi \alpha_{\rm eff}\right)
    \label{eq:An}
\end{equation}
and the best-fit parameter values are given in Table~\ref{tab:fit_params}. This function predicts $\Gamma \to 0$ as $\nu \to 0$, which is consistent with the expectation that infinitesimally small halos would have formed long time ago and not grow at all \citep[although they would still undergo pseudo-evolution,][]{diemer_13_pe}.

Figure~\ref{fig:all_fitting} demonstrates the performance of our universal fitting function (dashed lines). It matches the median MARs to 10\%, and often much better, across the full range of peak heights ($0.5 < \nu_{\rm 200m} < 5$), simulations (Erebos, \itng, \thesan), and redshift ($0 \leq z \leq 14$). The only exception are a few bins at low $\nu$ in \itng, which seem to be noisy. Importantly, the good agreement includes the self-similar cosmologies (bottom right panel of Figure~\ref{fig:all_fitting}), which present a stringent test of universality. 

Given that the model requires only six free parameters, the agreement constitutes a powerful demonstration of the notion that structure formation proceeds in a universal fashion when parameterized as a function of $\nu$, $n_{\rm eff}$, and $\alpha_{\rm eff}$. None of the $b_\rmi$ and $c_\rmi$ parameters in Table~\ref{tab:fit_params} are close to zero, indicating that both $n_{\rm eff}$ and $\alpha_{\rm eff}$ are necessary to describe both the normalization and $\nu$-dependence of the MAR. The dependence on the expansion history is more pronounced for MARs than for concentrations, where $\alpha_{\rm eff}$ has a noticeable impact only at the lowest redshifts \citep{diemer_19}.


\subsection{The distribution of MARs} \label{sec:res:distribution}

\begin{figure*}
    \centering
    \includegraphics[width=0.95\textwidth]{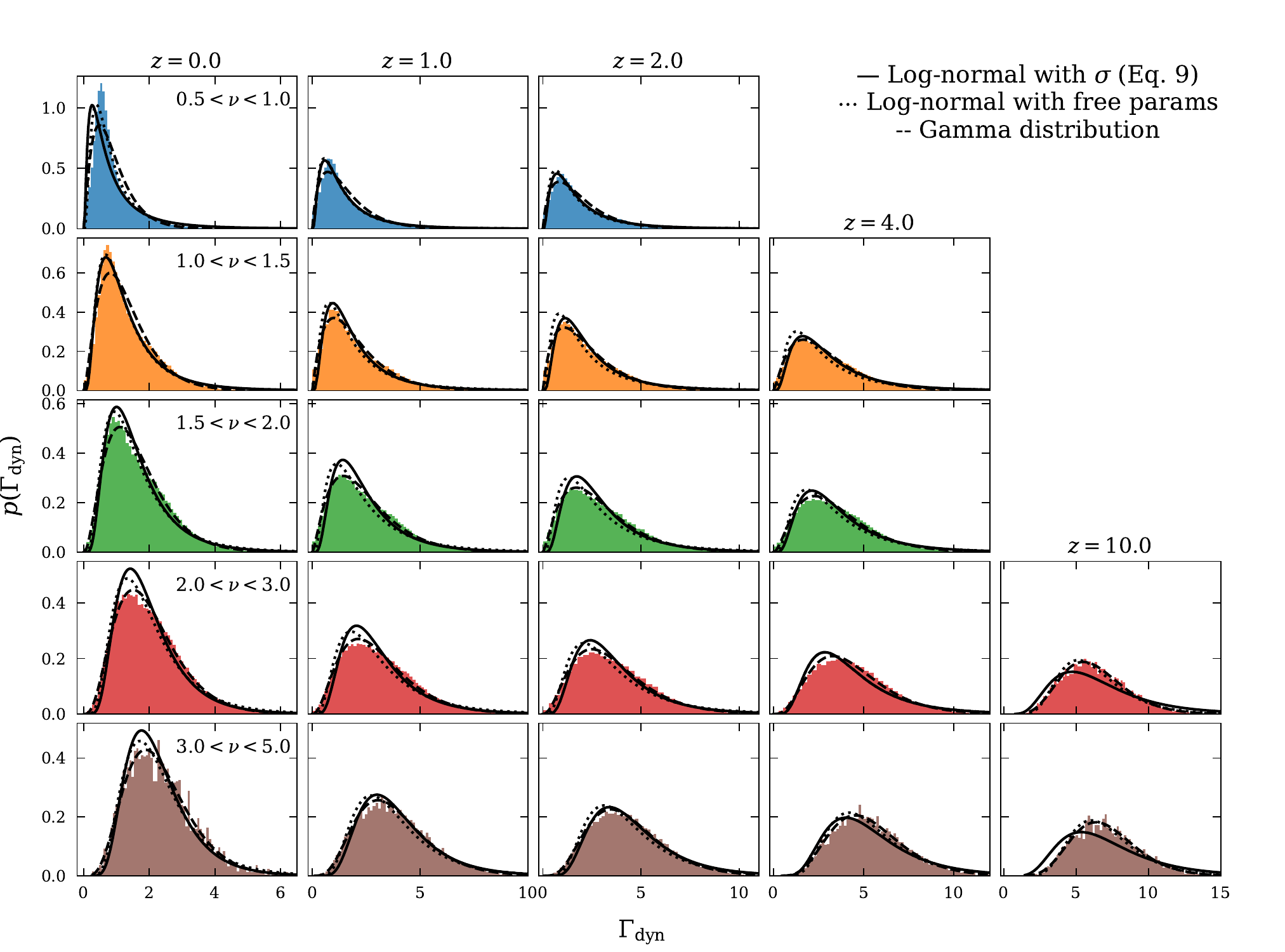}
    \caption{Distribution of $\Gamma_{\rm dyn}$ in bins of peak height $\nu$ (rows) and redshift (columns) for \erebos (WMAP7) simulation. The histograms are best described by the Gamma distribution fit with $f_{\rm loc}=0$ (dashed). However, a log-normal model specified entirely by the fitting functions of Equations~\ref{eq:fit_func} and~\ref{eq:scatter} (solid) also provides a reasonable fit, with KS statistics comparable to those of a log-normal fit with $\mu$ and $\sigma$ as free parameters (dotted).}
    \label{fig:gamma_distribution}
\end{figure*}

While the median MARs quantified in the previous section tell us about the important trends, there is also significant scatter in the distribution of $\Gamma_{\rm dyn}$, which is shown as a function of peak height and redshift in Figure~\ref{fig:gamma_distribution}. Since $\Gamma_{\rm dyn}$ is defined as a finite difference in $\log M$, it can be negative for halos that lose mass over a dynamical time due to tidal stripping, fly-by encounters, or transient misidentification by the halo finder. We find that the negative-$\Gamma$ fraction is confined to low $\nu$ and low $z$. For example, at $z=0$, $\sim 5\%$ of halos with $0.5 < \nu < 1.0$ have $\Gamma_{\rm dyn} < 0$, a fraction that drops to $< 1\%$ for $\nu > 1.0$. In the following, we therefore discard halos with $\Gamma_{\rm dyn} < 0$ and model only the distribution of positive accretion rates.

In particular, we characterize the scatter at fixed $\nu$ and $z$ using the log-normal width $\sigma(\ln\Gamma_{\rm dyn})$, defined as the half-width or $16$th--$84$th percentile interval. We find that this scatter is well approximated as 
\begin{equation}
    \sigma(\ln \Gamma_{\rm dyn}) = (s_0 + s_1\,\alpha_{\rm eff})\,\nu_{\rm 200m}^{-s_2}\,,
    \label{eq:scatter}
\end{equation}
where $s_0$, $s_1$, and $s_2$ are free parameters, and their best-fit values are given in Table~\ref{tab:fit_params}. In Figure~\ref{fig:gamma_distribution}, we compare three models for the MAR distribution, namely a Gamma distribution (dashed lines), a log-normal fit with free mean and scatter (dotted lines), and a log-normal where we fix the median to our best-fit value from Equation~\ref{eq:fit_func} and the log-normal scatter to Equation~\ref{eq:scatter} (solid lines). According to the Kolmogorov--Smirnov (KS) test \citep{massey_51}, the Gamma distribution provides the best fit with KS values of $\sim 0.01$--$0.05$ across nearly all $\nu$ and $z$ bins, while the log-normal with free parameters (dotted lines) performs comparably (KS $\sim 0.03$--$0.06$). However, given that the log-normal distribution with fixed parameters is much more constrained it still provides a reasonable fit (KS $\lesssim 0.05$ over most of the parameter space, rising to $\sim 0.09$--$0.13$ at $z = 0$ for low $\nu$ and at the highest redshifts).

\subsection{Formation time}
\label{sec:res:z_half}

\begin{figure*}
    \centering
    \includegraphics[trim = 2mm 2mm 2mm 2mm, clip, scale=0.75]{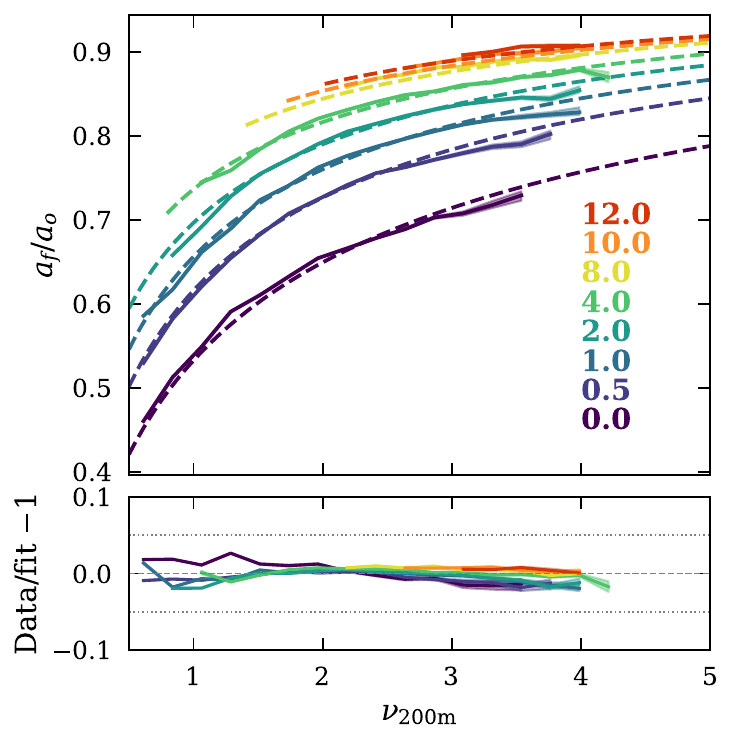}
    \includegraphics[trim = 19.5mm 2mm 2mm 2mm, clip, scale=0.75]{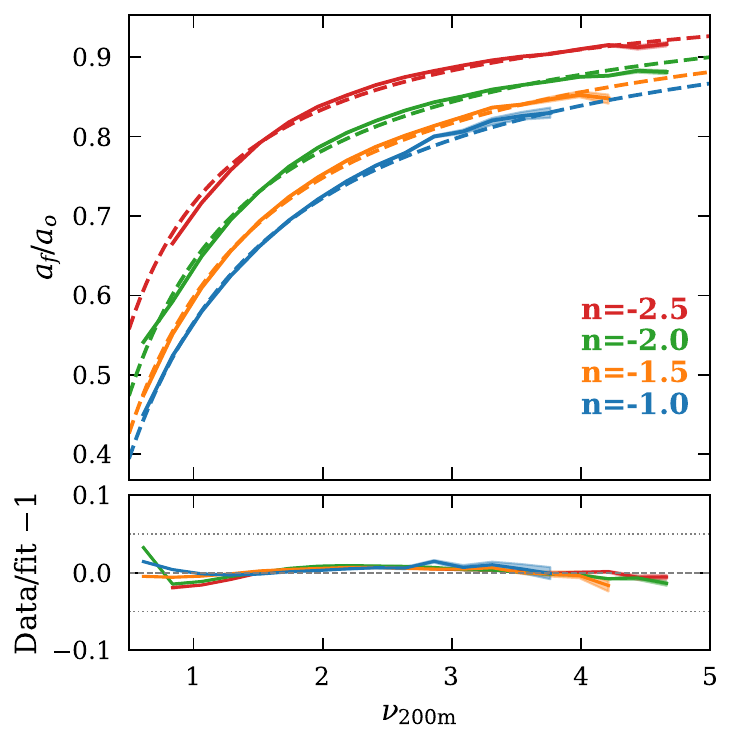}
    \caption{\textbf{Left:} Median ratio of the scale factor at formation to the scale factor at observation, $a_\rmf/a_\rmo$, as a function of peak height $\nu_{\rm 200m}$ for the $\Lambda$CDM simulations. Colors indicate the observation redshift, ranging from $z_0 = 0$ to $z_0 = 12$. Points show the binned medians measured from the simulation merger trees, and solid lines show the prediction from Equation~\ref{eq:af_lcdm} with $f = 0.5$ and $C(n_{\rm eff})$ given by Equation~\ref{eq:C_of_neff}. At low redshift, dark energy suppresses growth and shifts formation to earlier times (lower $a_\rmf/a_\rmo$), while at high redshift the curves converge as the universe becomes matter-dominated. \textbf{Right:} Same as left panel but for the self-similar simulations.  At fixed $\nu$, steeper power spectra (more negative $n$) produce later formation times.}
    \label{fig:zhalf}
\end{figure*}

A complementary characterization of halo growth is via models for the accretion history, which necessarily rely on a formation time. This time is often defined as the time (or redshift) $z_{1/2}$, the earliest epoch at which the most massive progenitor first exceeded half the final halo mass \citep{lacey_93}. Much past work has focused on the formation time of halos at $z = 0$, but we make no such restriction and express the formation time as $a_\rmf / a_\rmo$, the ratio of the scale factor where the halo attained half its mass relative to the mass at the scale factor of observation. Figure~\ref{fig:zhalf} shows the median ratios as a function of peak height and redshift for the WMAP7 $\Lambda$CDM and self-similar cosmologies. It approaches unity towards high redshift and peak height, which makes sense given the high values of $\Gamma_{\rm dyn}$ in this regime. As with the MARs, the differences between the self-similar simulations demonstrate that $a_\rmf/a_\rmo$ depends on $n$ at fixed $\nu$, with shallower spectra producing earlier formation times. 

In extended Press--Schechter (EPS) theory, the distribution of formation times can be expressed in terms of a dimensionless variable $\tilde{\omega}_\rmf = \Delta\delta_c(z) / \sqrt{\Delta S}$, where $\Delta\delta_c(z) = \delta_c(z_f) - \delta_c(z_\rmo)$ is the difference in the linear collapse threshold between the formation and observation epochs, and $\Delta S = \sigma^2(M_\rmo/2) - \sigma^2(M_\rmo)$ is the difference in the variance of the linear density field at the current mass $M_\rmo$ (i.e., the halo mass at the epoch of observation) and half that mass \citep{lacey_93}. The notation here means that $\delta_c(z) \equiv \delta_\rmc / D(z)$, and $\sigma$ is taken to refer to the linear density field at $z = 0$. \citet{lacey_93} thus write
\begin{equation}
\delta_c(z_f) = \delta_c(z_\rmo) + \tilde{\omega}_\rmf \sqrt{\sigma^2(fM_\rmo) - \sigma^2(M_\rmo)} \,.
\label{eq:zf_lc}
\end{equation}
Here $f$ is the mass fraction at the formation epoch, which should take on its physical value of $1/2$. We rewrite this equation as
\begin{equation}
    \frac{D(z_\rmf)}{D(z_\rmo)} = \left[ 1 + \frac{C}{\nu_\rmo} \,\sqrt{2\left(\frac{\sigma^2(fM_\rmo)}{\sigma^2(M_\rmo)} - 1\right)} \right]^{-1} \,,
    \label{eq:af_lcdm}
\end{equation}
where $\nu_\rmo$ signifies the peak height corresponding to $M_\rmo$ at $z_\rmo$ and $C \equiv \tilde{\omega}_\rmf / \sqrt{2}$ for historical reasons. In practice, we numerically solve this equation for $z_f$ to obtain $a_f / a_o = (1 + z_\rmo)/(1 + z_\rmf)$.

We now need to set the parameters $C$ and $f$. In EPS theory, the distribution of $\tilde{\omega}_\rmf$ can be predicted analytically and has a median of $\approx 0.974$ (obtained by solving Equation~2.33 of \citealt{lacey_93}), which is nearly independent of mass and spectral index. However, \citet{lacey_93} show that this prediction is not really borne out in simulations (their Figure~9). To make the formula agree better with simulations, \citet{vandenbosch_02_mah} introduced a calibration where $C = \mathrm{erf}^{-1}(1/2) \approx 0.477$ and $f = 0.254$, thus sticking to the physically expected value of $C$ and absorbing the difference between the EPS prediction and simulations into an effective formation fraction $f$ that differs from its physical value of $1/2$. 

We find that fitting for both $C$ and $f$ is ill-posed for reasons that we can understand by considering self-similar cosmologies. Here we have $D(a) = a$ and $\sigma(M) \propto M^{-(n+3)/6}$, which results in
\begin{equation}
\left. \frac{a_\rmf}{a_\rmo} \right|_{\rm self-sim} = \left[ 1 + \frac{C}{\nu_\rmo} \,\sqrt{2\left(f^{-(n+3)/3} - 1\right)} \right]^{-1} \,,
    \label{eq:af_selfsim}
\end{equation}
meaning that  $C$ and $f$ are entirely degenerate. In $\Lambda$CDM this degeneracy is weakly broken by the non-power-law shape of $\sigma(M)$, but in practice the simulation data do not constrain both parameters independently. We take the opposite approach to previous work and fix $f$ to its physical value of $f = 0.5$ while allowing $C$ to vary. Since the right panel of Figure~\ref{fig:zhalf} clearly shows that $C$ depends on $n$, we propose the simplest such modification possible,
\begin{equation}
    C(n_{\rm eff}) = C_0 + C_1\,n_{\rm eff} \,.
    \label{eq:C_of_neff}
\end{equation}
As for the parameters of our MAR formula, we fit $C_0$ and $C_1$ jointly over all self-similar simulations and all $\Lambda$CDM redshifts ($z_\rmo = 0 $--$ 12$) simultaneously. The best-fit parameters are given in Table~\ref{tab:fit_params}. Figure~\ref{fig:zhalf} demonstrates that our simple, two-parameter ansatz accurately captures the dependence of formation time on $n_{\rm eff}$, both in self-similar universes and in $\Lambda$CDM, with residuals at the $\lesssim 2\%$ level across the full range of $\nu_{200\rm m}$.

\section{Comparison with the literature} \label{sec:comp}

In this section we contrast our models for the MAR (Equation~\ref{eq:fit_func}) and the formation time (Equation~\ref{eq:af_lcdm}) to the literature. We begin with formation time models (Section~\ref{sec:comp:zform}) because those can be combined with a formula for the accretion history to predict MARs (Section~\ref{sec:comp:mar}).

\subsection{Models for the formation time}
\label{sec:comp:zform}

\begin{figure}
    \centering
    \includegraphics[width=\linewidth]{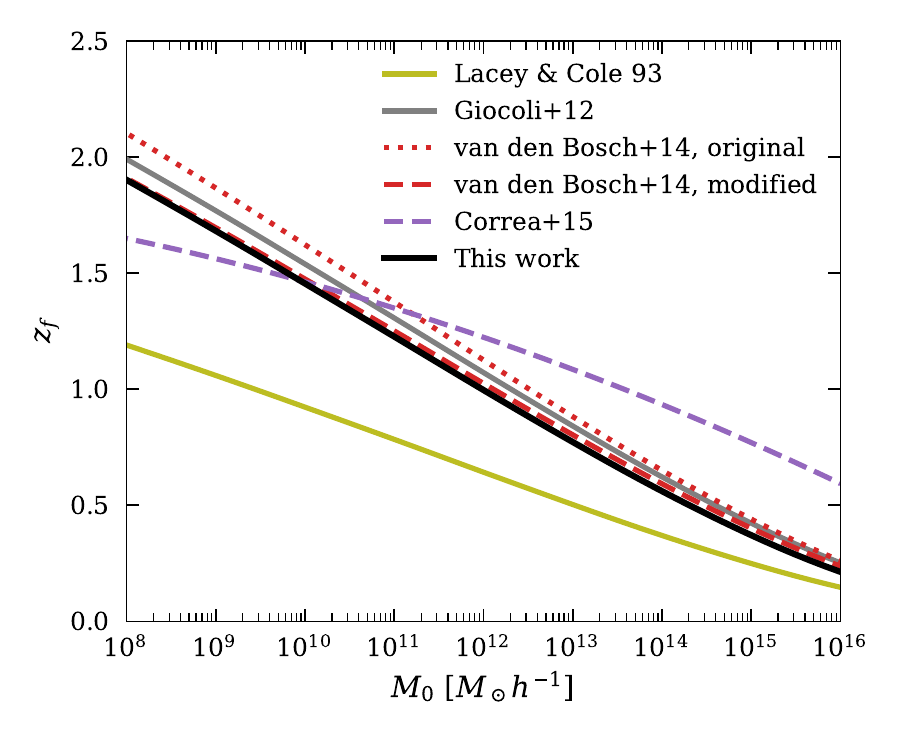}
    \caption{Median half-mass formation redshift $z_{1/2}$ as a function of halo mass at $z_0 = 0$ in the Planck cosmology. The solid black curve shows our work, obtained by solving Equation~\ref{eq:af_lcdm} with $f = 0.5$. The olive solid curve shows the \citet{lacey_93} EPS prediction (Equation~\ref{eq:zf_lc}) with $\tilde{\omega}_{\rmf} = \mathrm{erf}^{-1}(1/2)$ and $f = 0.5$. The red curves show \citet{vandenbosch_14} predictions with their original $\gamma = 0.4$ (dotted line) and our modification $\gamma = 0.2$ (dashed line) which brings it into excellent agreement with our fit. The gray solid curve shows \citet{giocoli_12} fit to their simulations with $f=0.5$. Finally, the purple dashed curve shows the fitting formula of \citet{correa_15_a}.}
    \label{fig:zf_comparison}
\end{figure}

Figure.~\ref{fig:zf_comparison} compares our model for the half-mass redshift for halos at $z = 0$ to models from the literature. The EPS-based prediction of \citet[][olive solid line]{lacey_93} lies systematically below all other models, reproducing a well-known underestimate of formation times \citep[][see also the discussion in Section~\ref{sec:res:z_half}]{giocoli_07}. However, our model matches well with those of \citet[][gray solid line]{giocoli_12} and \citet[][red dotted line]{vandenbosch_14}, despite the three models being calibrated on different simulation suites and cosmologies. The match to \citet{vandenbosch_14} can even be improved by making a modification to one of their free parameters (red dashed curve), which we describe in detail in the following section. Finally, we compare to the model of \citet[][purple dashed line]{correa_15_a}, which predicts a much weaker dependence on halo mass and agrees with the other models only around $M \sim 10^{11}\,h^{-1}M_\odot$.

\subsection{Models for the MAR}
\label{sec:comp:mar}

\begin{figure*}
    \centering
    \includegraphics[trim = 2mm 18.5mm 2mm 2mm, clip, scale=0.5]{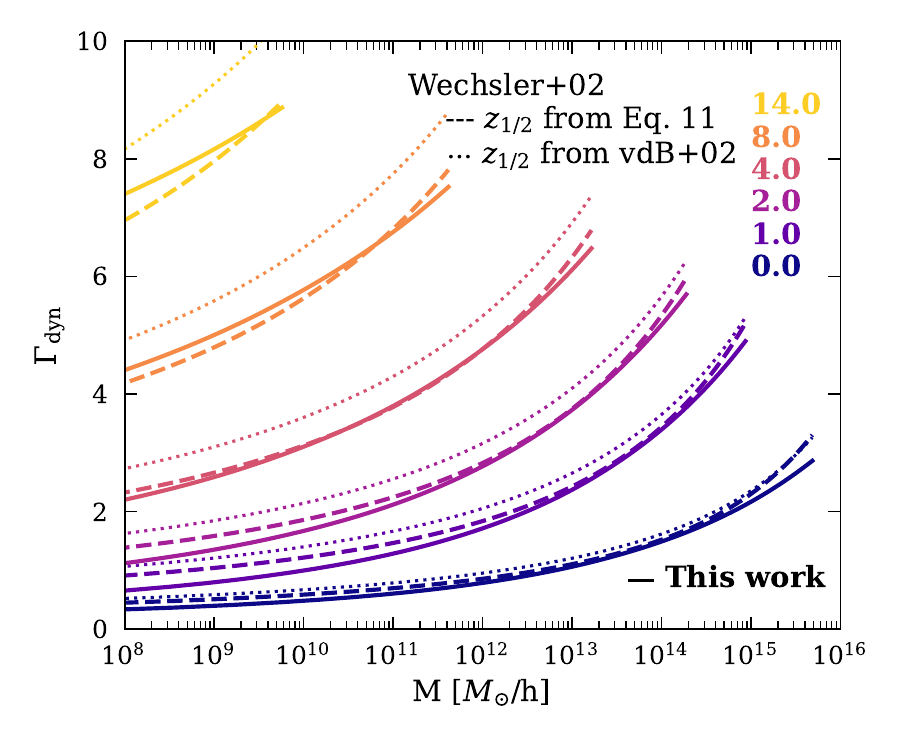}
    \includegraphics[trim = 18mm 18.5mm 2mm 2mm, clip, scale=0.5]{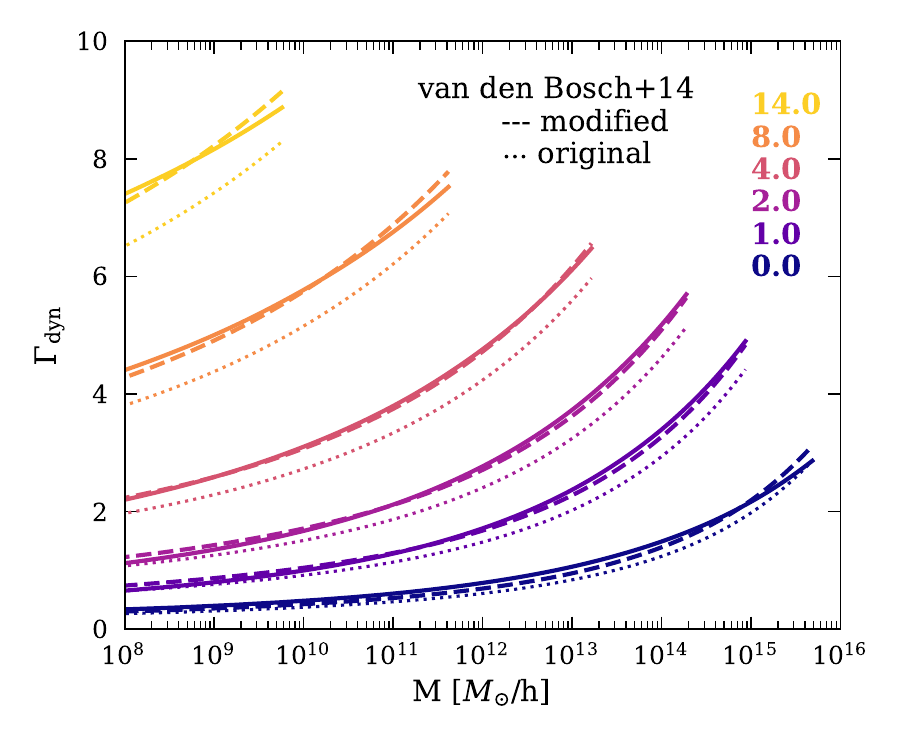}
    \includegraphics[trim = 2mm 18.5mm 2mm 2mm, clip, scale=0.5]{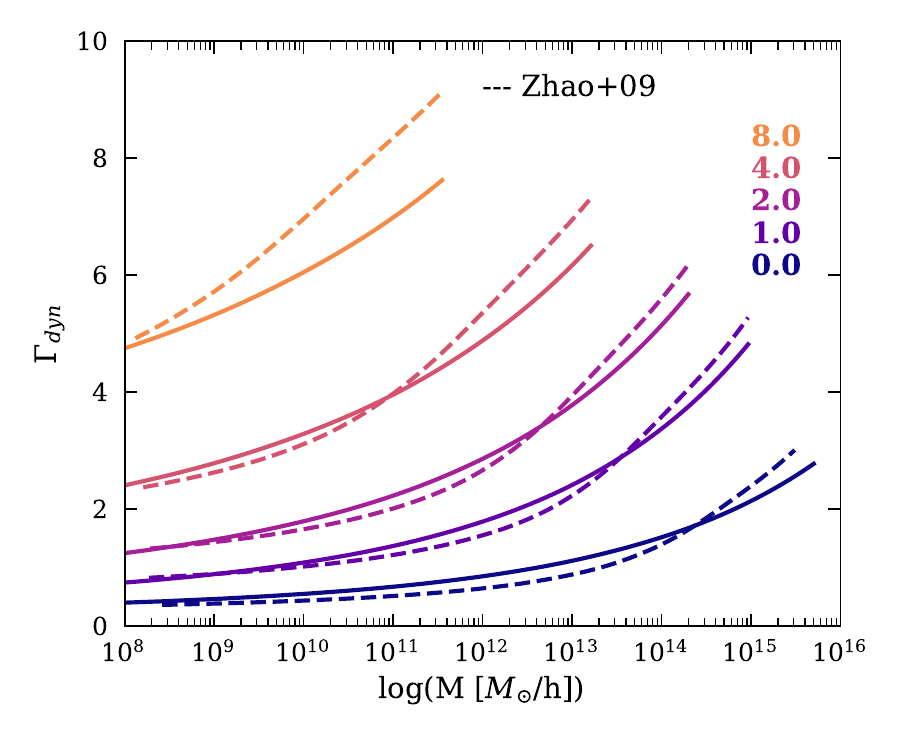}
    \includegraphics[trim = 18mm 18.5mm 2mm 2mm, clip, scale=0.5]{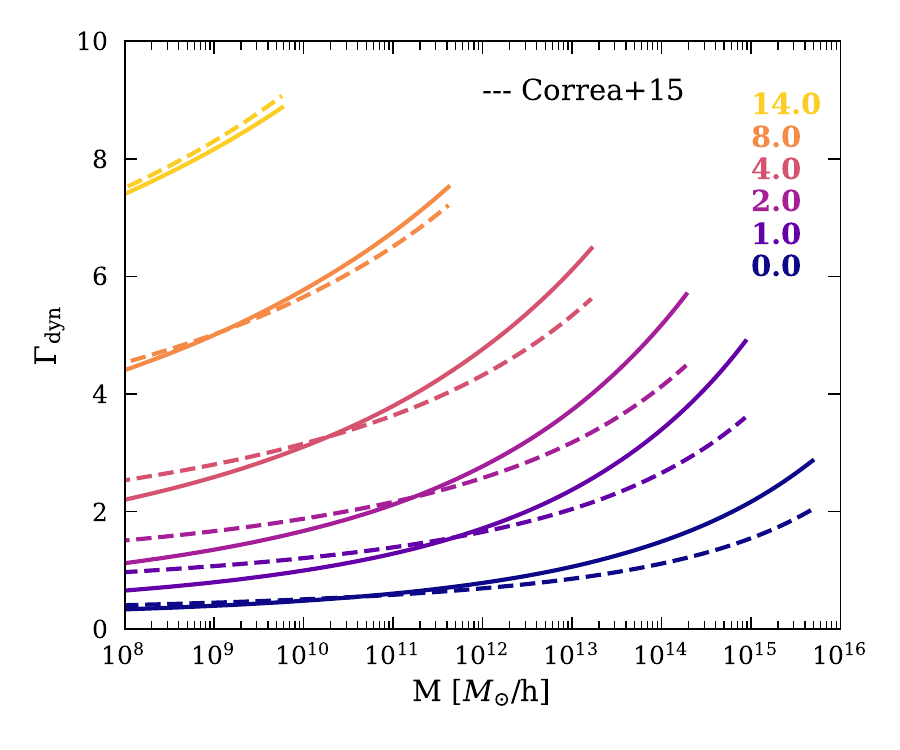}
    \includegraphics[trim = 2mm 2mm 2mm 2mm, clip, scale=0.5]{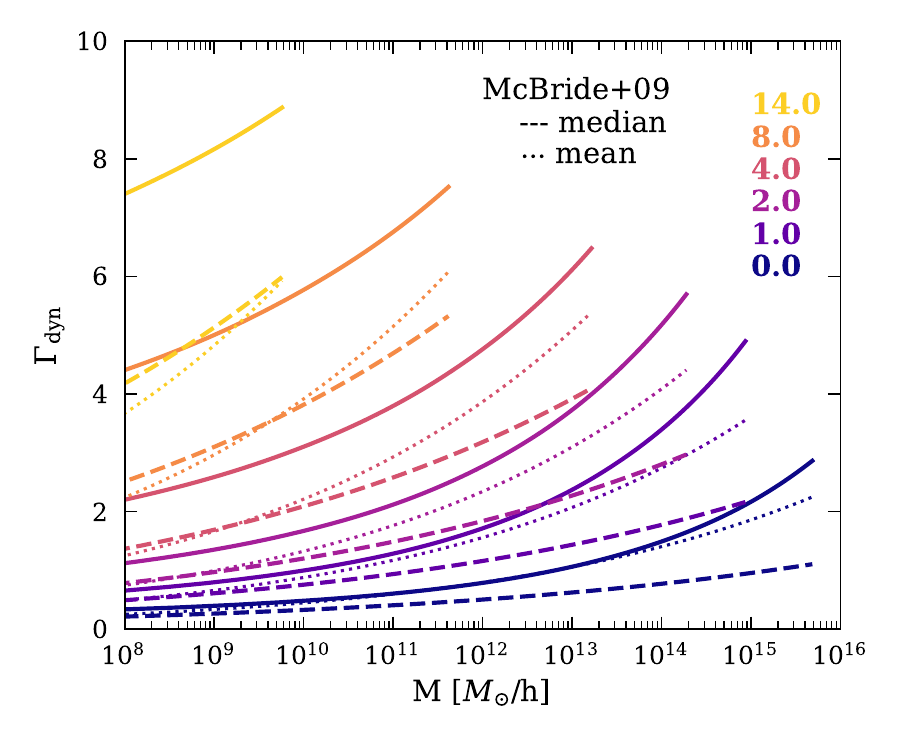}
    \includegraphics[trim = 18.5mm 2mm 2mm 2mm, clip, scale=0.5]{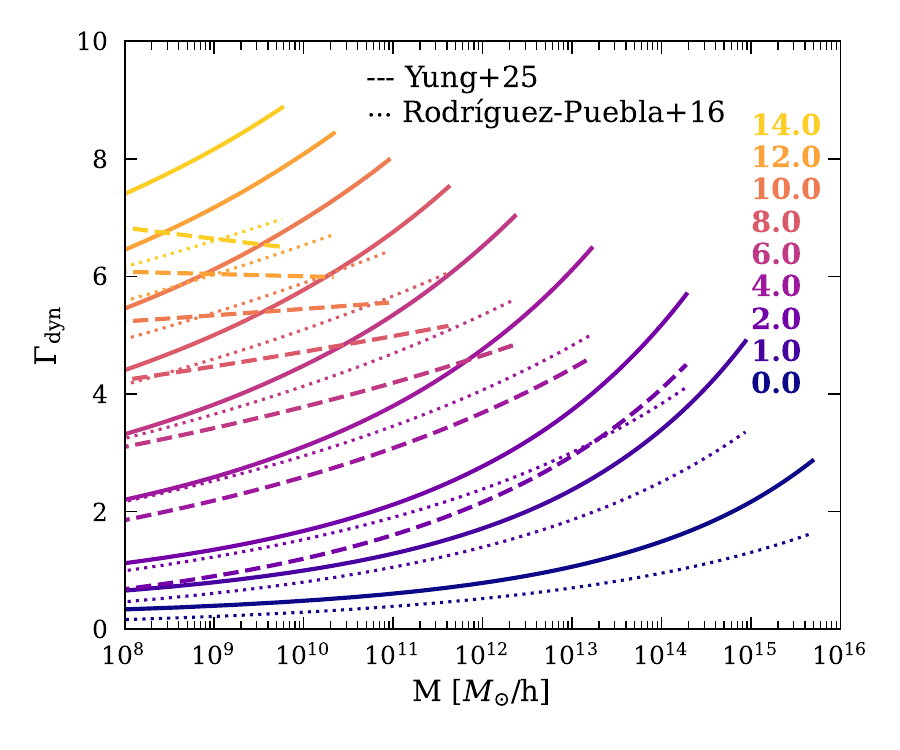}
    \caption{Comparison of $\Gamma_{\rm dyn}(M, z)$ from our fitting function (solid lines) with models from the literature, with colors indicating redshift from $z = 0$ (dark blue) to $z = 14$ (yellow). All models have been converted to $\Gamma_{\rm dyn}$ measured over one dynamical time (Section~\ref{sec:comp:mar}). \textbf{Top left:} the exponential MAH of \citet{wechsler_02} with formation redshifts based on our model (dashed) and the \citet{vandenbosch_02_mah} calibration (dotted). 
    \textbf{Top right:} the \citet{vandenbosch_14} model for the median (dotted), as well as  an updated version with $\gamma = 0.2$ (Section~\ref{sec:comp:mar:z12}). 
    \textbf{Middle left:} the \citet{zhao_03_mah} two-phase MAH model, evaluated for the Bolshoi cosmology. 
    \textbf{Middle right:} the \citet{correa_15_c} semi-analytic MAH model, which agrees with our fit at high redshift ($z \gtrsim 4$) but under-predicts $\Gamma_{\rm dyn}$ at lower redshifts for massive halos due to the differences in the formation redshift (see Figure~\ref{fig:zf_comparison}).
    \textbf{Bottom left:} the \citet{mcbride_09} model for $\dot{M}(M,z)$, showing both the median (dashed) and mean (dotted). 
    \textbf{Bottom right:}: the \citet{rodriguezpuebla_16} (dotted) and \citet{yung_25} (dashed; $z \geq 2$) predictions.}
    \label{fig:all_comparison}
\end{figure*}

There are fundamentally two ways to model MARs: directly as in this paper, or indirectly by combining a model for formation times with a formula for accretion histories. Figure~\ref{fig:all_comparison} presents a comparison between our universal fitting function and several previously proposed models of each type. Unlike Figure~\ref{fig:all_fitting}, we show $\Gamma_{\rm dyn}$ as a function of halo mass to facilitate comparison with earlier work. In each panel, solid lines show our model and dashed or dotted lines show the comparison model, with colors indicating redshift from $z = 0$ (dark blue) to $z = 14$ (yellow). Unless otherwise mentioned, all comparisons are performed in the \citet{2020A&A...641A...6P} cosmology. To compare the numerous different definitions of the MAR on equal footing, we convert all models to our definition of $\Gamma_{\rm dyn}$, i.e. the relative mass change in $M_{\rm 200m}$ over a crossing time, as described below.

\subsubsection{Models based on formation times and MAHs}
\label{sec:comp:mar:z12}

The models discussed in this section provide a description of the full MAH $M(z)$. We evaluate those functions at the redshift of observation $z_{\rm o}$ and one dynamical time earlier to take the finite difference.

The \citet{wechsler_02} exponential MAH model, $M(z) = M_0\exp(-\alpha z)$, provides a one-parameter description of halo growth where the rate parameter $\alpha$ is related to the formation redshift. We calculate $\alpha$ in two different ways. First, we use Equation~\ref{eq:zf_lc} and convert $\alpha = \ln 2 / (z_f - z_o)$, which follows directly from the definition of the half-mass redshift (dashed lines in top-left panel of Figure~\ref{fig:all_comparison}). The agreement with our model (solid lines) is remarkable given the simplicity of the underlying exponential MAH. This comparison demonstrates that the simple exponential form remains accurate at high redshift when anchored by our $C(n_{\rm eff})$ calibration of the formation time.
Second, we calculate $\alpha$ based on the \citet{vandenbosch_02_mah} calibration of formation time ($f = 0.254, C = 0.477$) as well as their suggested conversion, $\alpha = (z_f \times 1.43)^{-1.05}$ (dotted lines). This version is close to our model at $z = 0$ but diverges progressively at higher redshifts, which is expected since the \citet{vandenbosch_02_mah} formation times were calibrated only at $z = 0$.

\citet{vandenbosch_14} presented an updated, universal MAH model calibrated to the Bolshoi simulation \citep{klypin_11}. In this model, the progenitor mass fraction $\psi = M(z)/M_0$ is expressed via an implicit function $F(\psi) = \tilde{\omega}$, where $\tilde{\omega}$ is the EPS-based rescaled time variable. We invert the equation to obtain $\psi$ and take a finite difference to calculate $\Gamma_{\rm dyn}$. The results are close to our model at $z \lsim 2$ but predict lower MARs at higher redshift (dotted lines in top-right panel of Figure~\ref{fig:all_comparison}). 

However, we found that a small parameter change in the \citet{vandenbosch_14} model can bring our models into remarkably close agreement. In particular, the rescaled time variable, which accounts for EPS barrier crossing statistics, involves a correction term $G^{\gamma}$, where $\gamma$ controls the strength of the \citet{parkinson_08} modification to the standard EPS barrier. The original calibration of \citet{vandenbosch_14} adopted $\gamma = 0.4$ based on fits to the Bolshoi simulation at $z \lesssim 2$, with EPS-based merger trees extending to higher redshifts. Reducing this parameter to $\gamma = 0.2$ brings our models into agreement (dashed lines in top-right panel of Figure~\ref{fig:all_comparison}). This match is in line with the almost perfect agreement of the formation time that we found in Section~\ref{sec:comp:zform} and Figure~\ref{fig:zf_comparison}. We note that the value of $\gamma = 0.2$ was found by experimentation rather than through a formal refit to the simulation data, and that no other parameters were modified. This suggests that a formal recalibration of the correction term against high-redshift simulation data could yield a particularly robust and universal MAH model. Regardless, the agreement of our model with \citet{giocoli_12} and \citet{vandenbosch_14} demonstrates that our $C(n_{\rm eff})$ parametrization captures similar physics as the empirical correction to standard EPS encoded in the $G^\gamma$ term, namely, the dependence of halo assembly on the local slope of the power spectrum, which is not captured by a constant-barrier, Markovian EPS formalism.

\citet{zhao_09_acchist} proposed a two-phase model where the MAH experiences a fast-accretion phase dominated by major mergers, followed by a slow phase of minor accretion (and pseudo-evolution). We evaluate their MAH model using a public code provided by the authors. The agreement is good at $z \lesssim 4$, but the model over-predicts our MARs at higher redshifts ($z\sim 8$), with the discrepancy growing toward higher masses (middle-right panel of Figure~\ref{fig:all_comparison}).

Additionally, we compare to the semi-analytical MAH model of \citet{correa_15_c}, which is defined as $\tilde{M}(z, M(z_i), z_i) = M(z_i)(1+z-z_i)^{\tilde{\alpha}}\,e^{\tilde{\beta}(z-z_i)}$, where $\tilde{\alpha}$ and $\tilde{\beta}$ depend on halo mass and cosmology (through $\sigma(M)$ and the linear growth factor) as well as on $z_f$. The comparison (bottom-right panel of Figure~\ref{fig:all_comparison}) shows good agreement at high redshift ($z \gtrsim 4$). At lower redshifts, however, the \citet{correa_15_c} model systematically under-predicts our MARs. This discrepancy can be traced to differences in the formation redshift as shown in Figure~\ref{fig:zf_comparison}, where the \citet{correa_15_c} $z_f(M)$ relation is significantly shallower than ours and thus predicts earlier formation for massive halos ($z_f \approx 0.6$ vs.\ our $\approx 0.2$ at $M \sim 10^{15}\,\msunh$), which translates into lower mass accretion later. 

We have also compared our function to the DiffMAH model of \citet{hearin_21} by creating MAHs with their population model (DiffMAHpop) and taking the median. At $z = 0$, the two models agree well at intermediate masses ($M \sim 10^{12} $--$ 10^{14}\,\msunh$), but DiffMAHpop predicts smaller $\Gamma_{\rm dyn}$ by up to a factor of $\sim 2$ at low and high masses. DiffMAHpop is currently calibrated only for halos identified at $z = 0$, and re-fitting the DiffMAH parameters at higher redshift is beyond the scope of this work.

\subsubsection{Models that directly parameterize the MAR}
\label{sec:comp:mar:other}

The models discussed in this section provide some measure of the instantaneous MAR, $\dot{M}(M, z)$, which we convert to $\Gamma_{\rm dyn}$ by integrating the mass growth backward over one dynamical time, accounting for redshift and mass-dependent changes in the MAR formulae.

\citet{mcbride_09} parameterized the instantaneous MAR as a power law in mass with redshift-dependent normalization, providing separate fits for the mean and median MAR,
\begin{equation}
    \frac{\left\langle \dot{M} \right\rangle}{[M_\odot\,{\rm yr}^{-1}]} = 42.0 \left(\frac{M}{10^{12}\,M_\odot}\right)^{1.127} (1 + 1.17\,z)\,E(z)
\end{equation}
and
\begin{equation}
    \frac{\dot{M}_{\rm med}}{[M_\odot\,{\rm yr}^{-1}]} = 24.1 \left(\frac{M}{10^{12}\,M_\odot}\right)^{1.094} (1 + 1.75\,z)\,E(z) \,,
\end{equation}
where $E(z) = \sqrt{\Omega_m(1+z)^3 + \Omega_\Lambda}$ and $M$ is in solar masses. The middle-left panel of Figure~\ref{fig:all_comparison} shows that this model agrees with ours at $z \lesssim 1$, with the mean relation (dotted lines) within $\sim 10 $--$ 20\%$, whereas the median MAR (dashed lines) is systematically lower at all redshifts. At higher redshifts, both the mean and median increasingly under-predict $\Gamma_{\rm dyn}$ relative to our fit. This divergence may be due to limitations in calibration range, since the \citet{mcbride_09} function was fitted to the Millennium simulation at $z \leq 6$.

In the bottom-left panel of Figure~\ref{fig:all_comparison}, we compare our fitting function with those of \citet[][dotted]{rodriguezpuebla_16} and \citet[][dashed]{yung_25}, which follow a similar mathematical formalism but were trained on different simulations and redshifts. To ensure a fair comparison, all three models are evaluated for the \citet{2016A&A...594A..13P} cosmology. We additionally convert $\Gamma_{\rm dyn}$ to the different dynamical time definition used for the comparison models, namely half the crossing time, $t_{\rm peri} = R_{\rm vir}/V_{\rm vir}$. To convert from the given accretion rate, we note that $\Gamma_{\rm dyn}$ as defined in Eq.~\ref{eq:gamma_def} can be written in as a mass ratio,
\begin{equation}
\frac{M(t)}{M(t - t_{\rm dyn})} = \left(\frac{a(t)}{a(t - t_{\rm dyn})}\right)^{\Gamma_{\rm dyn}}\,.
\label{eq:mass_ratio}
\end{equation}
We now substitute into the definition of the dynamical-time-averaged specific accretion rate according to \citet{rodriguezpuebla_16} while maintaining the meaning of $\Gamma_{\rm dyn}$,
\begin{equation}
\frac{\langle \dot{M} \rangle_{\rm peri}}{M(t)} = \frac{1}{t_{\rm peri}} \left[1 - \left(\frac{a(t - t_{\rm peri})}{a(t)}\right)^{\Gamma_{\rm dyn}}\right]\,,
\label{eq:smar_from_gamma}
\end{equation}
which can be inverted to give
\begin{equation}
\Gamma_{\rm dyn} = \frac{\ln \left[1 - \frac{\langle \dot{M} \rangle_{\rm peri}}{M(t)}\, t_{\rm peri}\right]}{\ln \left[ a(t - t_{\rm peri})/a(t) \right]}\,.
\label{eq:gamma_from_smar}
\end{equation}
At $z \lesssim 6$, the \citet{rodriguezpuebla_16} model predicts MARs that are systematically lower by $\sim 0.2$--$0.5$ across most of the mass range ($10^{11} - 10^{16}\, \msunh$). At higher redshifts, this discrepancy grows considerably, reaching $\Delta\Gamma \sim 1$--$2$ by $z \sim 14$. This divergence is largely an artifact of the nonlinear mapping between the specific accretion rate and $\Gamma_{\rm dyn}$: the denominator of Equation~\ref{eq:gamma_from_smar} decreases with increasing redshift, meaning that even a $\sim 20\%$ difference in $\langle \dot{M} \rangle / M$ is amplified into a much larger offset in $\Gamma_{\rm dyn}$. 

The model of \citet{yung_25}, which is calibrated on the Bolshoi-Planck/MultiDark and \gureft simulation suites, predicts quite different $\Gamma_{\rm dyn}$ than all other models at $z> 6$. Specifically, it underestimates $\Gamma_{\rm dyn}$ relative to our fit at $z \leq 6$ and then matches at lower masses for $z \gtrsim 8$, while simultaneously showing a turnover where $\Gamma_{\rm dyn}$ decreases toward higher masses. This behavior arises from the model formulation, $\dot{M} \propto [M\,E(z)]^{\alpha}$, which means that $\dot{M}/M \propto M^{\alpha - 1}$. The exponent $\alpha$ drops from $\approx 1.1$ at $z = 2$ to below unity at $z \gtrsim 12$. With $\alpha < 1$, the model thus becomes a decreasing function of mass, which is opposite to the expected hierarchical growth. Even at $z = 10$ where $\alpha$ is only marginally above unity, the nearly mass-independent specific rate produces a much flatter $\Gamma_{\rm dyn}(M)$ relation than our fit, underestimating the accretion rates of the most massive halos. This may reflect the limited mass range of the \gureft suite at high redshift, as discussed in \citet{yung_25}.

In summary, our fitting function is broadly consistent with previous models in the regimes where they were calibrated, but no existing model reproduces our MARs over their full dynamic range. The key advantage of our approach is its universality, as demonstrated by its fit to extreme, self-similar cosmologies.

\section{Conclusions} \label{sec:conclude}

We have presented a universal model for the median mass accretion rates and formation times of dark matter halos. These models are calibrated on a large sample of halos ($\sim$ over $12$ million) drawn from the \erebos, \itng, and \thesan simulation suites, which include $\Lambda$CDM and self-similar Einstein-de Sitter cosmologies. Our main results are as follows.

\begin{itemize}

\item The median MARs in simulations are remarkably robust to methodological differences such as simulation codes, halo finders (Rockstar vs. Subfind), merger-tree algorithms (SubLink vs. LHaloTree), and resolution. 

\item Baryonic physics has essentially no impact on the median mass accretion rates measured across $R_{200\rm m}$, although we caution that this conclusion may not extend to much smaller radii such as $R_{500c}$, where feedback can redistribute gas more effectively.

\item The median MAR averaged over a dynamical time, $\Gamma_{\rm dyn}$, can be expressed as a universal function of three quantities: the peak height $\nu$, the effective slope of the linear power spectrum $n_{\rm eff}$, and the effective exponent of linear growth $\alpha_{\rm eff}$. The resulting fitting function (Equation~\ref{eq:fit_func}) requires only six free parameters to achieve better than 10\% accuracy across $\Lambda$CDM cosmologies from $z = 0$ to $z = 14$, as well as self-similar cosmologies with power-law spectra from $n = -1$ to $-2.5$. 

\item We derive a complementary fitting function for the half-mass formation redshift $z_{1/2}$ by revisiting the formalism of \citet{lacey_93}. Within this framework, we fix the parameter $f$ to is physical value of $1/2$ and introduce a normalization $C$ that depends on $n_{\rm eff}$ (Equation~\ref{eq:C_of_neff}), leading to an accurate model with only two free parameters.

\item Combining our function for $z_{1/2}$ with the accretion history model of \citet{wechsler_02}, we obtain MARs that agree with our direct fit to within $\sim 10$--$20\%$, demonstrating the internal consistency of our framework. 

\item While MAR models from the literature broadly agree with our model where they were calibrated, no previous model matches our function (and thus our simulation data) for all cosmologies, masses, and redshifts. Some models can be brought intro agreement either by using our model for formation time or with tweaks to certain free parameters.


\end{itemize}

Our model has immediate practical applications. Given only a halo's current mass and redshift, one can estimate both the mass accretion rate (from the $\Gamma_{\rm dyn}$ fitting function) and the mass assembly history (using the formation redshift), without recourse to merger-tree information. This capability is valuable for any model that ties star formation to halo growth \citep[e.g.,][]{sun_16, mirocha_17, park_19}, and for observational studies where accretion rates must be inferred from instantaneous halo properties. Our predictions extend to $z \sim 14$, precisely the regime now probed by JWST observations of early galaxy formation. Coupling the MAR and $z_{1/2}$ fits to galaxy formation models could yield direct predictions for high-redshift stellar mass growth and UV luminosity functions.

Extending this calibration to non-standard scenarios such as $w$CDM, modified gravity, or warm/self-interacting dark matter would test the robustness of the parameterization beyond standard structure formation, which we defer to future work. Additionally, decomposing $\Gamma_{\rm dyn}$ into major-merger, minor-merger, and smooth-accretion contributions would provide a more detailed picture relevant to galaxy morphology, since these channels are thought to drive distinct evolutionary pathways \citep{toomre_72, keres_05, dekel_09}. Such a decomposition has been performed in individual simulations \citep[e.g.,][]{genel_10, fakhouri_10}, though the distinction between minor mergers and smooth accretion is inherently resolution-dependent. Extending this decomposition within our universal framework would test whether the different accretion channels are themselves captured by the same set of variables. 

The fitting functions presented here are implemented in the public \textsc{Colossus} toolkit \citep{diemer_18_colossus}.


\begin{acknowledgments}

This material is based upon work supported by the U.S. National Science Foundation under Grant No. NSF AST 2338388. We thank Frank van den Bosch, Andrew Hearin, and Aaron Yung for useful discussions. AB acknowledges the financial support from the CTC fellowship at the University of Maryland. The computations were run on the Zaratan cluster at the University of Maryland. 

\end{acknowledgments}


\software{Colossus \citep{diemer_18_colossus}, Astropy \citep{2013A&A...558A..33A,2018AJ....156..123A,2022ApJ...935..167A}, Matplotlib \citep{hunter_matplotlib}, Numpy \citep{harris_numpy}, SciPy \citep{virtanen_scipy}}

\bibliography{bib/bib_general, bib/bib_structure, bib/bib_simulations}{}
\bibliographystyle{aasjournalv7}

\end{document}

%% file: macros_general.tex
\def\gtsima{$\; \buildrel > \over \sim \;$}
\def\ltsima{$\; \buildrel < \over \sim \;$}
\def\prosima{$\; \buildrel \propto \over \sim \;$}
\def\gsim{\lower.7ex\hbox{\gtsima}}
\def\lsim{\lower.7ex\hbox{\ltsima}}
\def\simgt{\lower.7ex\hbox{\gtsima}}
\def\simlt{\lower.7ex\hbox{\ltsima}}
\def\simpr{\lower.7ex\hbox{\prosima}}

\newcommand{\kpch}{{h^{-1}{\rm kpc}}}
\newcommand{\mpch}{h^{-1}{\rm {Mpc}}}

\newcommand{\msunh}{h^{-1} M_\odot}

\def\erebos{Erebos\xspace}

\def\itng{IllustrisTNG\xspace}

\def\thesan{Thesan\xspace}
\def\gureft{GUREFT\xspace}

\def\rmc{{\rm c}}

\def\rmf{{\rm f}}

\def\rmi{{\rm i}}

\def\rmo{{\rm o}}

\def\bm{{\bf m}}